\documentclass[twocolumn,pre,longbibliography,superscriptaddress]{revtex4-1}
\usepackage{natbib}
\usepackage[utf8]{inputenc}
\usepackage{amsmath,amssymb,color}
\usepackage{graphicx}
\usepackage{latexsym}
\usepackage{multirow}

\newcommand{\textin}[1]{\mbox{\scriptsize{#1}}}

\begin{document}

\title{Effect of an axial electric field on the breakup of a leaky-dielectric liquid filament}
\author{M. Rubio}
\address{Departmento de Ingenier\'{\i}a Mec\'anica, Energ\'etica y de los Materiales and\\
Instituto de Computaci\'on Cient\'{\i}fica Avanzada (ICCAEx),\\
Universidad de Extremadura, Avda.\ de Elvas s/n, E-06071 Badajoz, Spain}
\author{A. Ponce-Torres}
\address{Departmento de Ingenier\'{\i}a Mec\'anica, Energ\'etica y de los Materiales and\\
Instituto de Computaci\'on Cient\'{\i}fica Avanzada (ICCAEx),\\
Universidad de Extremadura, Avda.\ de Elvas s/n, E-06071 Badajoz, Spain}
\author{M. A. Herrada}
\address{Departamento de Mec\'anica de Fluidos e Ingenier\'{\i}a Aeroespacial,\\
Universidad de Sevilla, E-41092 Sevilla, Spain}
\author{A. M. Gañán-Calvo}
\address{Departamento de Mec\'anica de Fluidos e Ingenier\'{\i}a Aeroespacial,\\
and Laboratory of Engineering for Energy and Environmental Sustainability,\\
Universidad de Sevilla, E-41092 Sevilla, Spain}
\author{J. M. Montanero}
\address{Departmento de Ingenier\'{\i}a Mec\'anica, Energ\'etica y de los Materiales and\\
Instituto de Computaci\'on Cient\'{\i}fica Avanzada (ICCAEx),\\
Universidad de Extremadura, Avda.\ de Elvas s/n, E-06071 Badajoz, Spain}

\begin{abstract}
We study experimentally and numerically the thinning of a Newtonian leaky-dielectric filament subject to an axial electric field. We consider moderately viscous liquids with high permittivity. The experiments show that satellite droplets are produced from the breakup of the filaments with high electrical permittivity due to the action of the electric field. Two electrified filaments with the same minimum radius thin at the same speed regardless of when the voltage was applied. The numerical simulations show that the polarization stress is responsible for the pinching delay observed in the experiments. Asymptotically close to the pinching point, the filament pinching is dominated by the diverging hydrodynamic forces. The polarization stress becomes subdominant even if this stress also diverges at this finite-time singularity.
\end{abstract}


\maketitle

\section{Introduction}

The pinching of a liquid's free surface is a spontaneous and ubiquitous phenomenon that allows us to observe the behavior of fluids with length and time scales decreasing down to the breakdown of the continuum hypothesis \citep{EV08}. The increasing surface-to-volume ratio in the pinching point unveils interfacial effects obscured in experiments with much larger spatial dimensions. The decreasing time scale of this phenomenon can reveal physical properties of the fluid not observable under normal conditions \citep{RPVHM19}.

Very close to the pinching, the spatial and time scales are small enough for the system to adopt a universal behavior, independent from the boundary and initial conditions, and affected only by all or some of its intrinsic properties (density $\rho$, viscosity $\mu$ and surface tension $\gamma$). The thinning of low-viscosity fluid threads passes through an inertio-capillary regime in which the free surface minimum radius, $R_{\textin{min}}$, and the characteristic axial length of the pinching region, $\ell$, obey the power laws 
\begin{equation}
\label{inv}
R_{\textin{min}}\sim \ell \sim \tau^{2/3}, \quad 
\end{equation}
where $\tau$ is the time to the pinching \citep{KM83}. The breakup of high-viscosity liquid filaments goes through the viscous-capillary mode \citep{P95} in which
\begin{equation}
\label{vis}
R_{\textin{min}}(\tau)=0.0709\  \gamma/\mu\ \tau, \quad \ell\sim \tau^{0.175}.
\end{equation}
Sufficiently close to the pinching point, the system adopts a final inertio-viscous-capillary regime in which all three forces are commensurate with each other, and the free surface minimum radius obeys the asymptotic law \citep{E93}
\begin{equation}
\label{uni}
R_{\textin{min}}(\tau)=0.0304\  \gamma/\mu\ \tau, \quad \ell\sim \tau^{1/2}.
\end{equation}
Several intermediate transitions between the three regimes mentioned above can occur before the ultimate inertial-viscous-capillary regime is reached \citep{CCTSHHLB15,LS16}.

Electric fields are frequently applied to power the ejection and control the breakup of liquid threads in many microfluidic applications \citep{T64,DAMHL03,CJHB08,V19,MG20}. Researchers have addressed how radial electric fields affect the dynamics of conductor threads close to the interface pinching. \citet{CHB07} integrated the full Navier-Stokes equations to describe the breakup of a perfectly conductor cylindrical jet immersed in a radial electric field. Their results show that the nonlinear terms generally delay the jet breakup. The diameter of the satellite droplet increases with the electric strength, especially for large values of the Ohnesorge number. Electrostatic interfacial stresses produce similar effects to those of inertia, leading to the formation of satellite droplets even in the Stokes limit \citep{CHB07}. For low values of the Ohnesorge number, the electric field enhances the free surface overturning, which results in total shielding of the pinch region from the radial electric field. As a consequence, the electric field has little influence on the local pinch-off dynamics. In the Stokes limit, the behavior asymptotically close to the pinching might be affected by the electric stresses because the overturning does not occur in this case. However, \citet{WP11} found that the pinch-off dynamics of a viscous thread surrounded by a viscous bath are dominated by hydrodynamic forces. In fact, the electric to capillary pressure ratio becomes asymptotically small at pinching. For this reason, the asymptotic solution is self-similar and recovers that of the non-electrified case \citep{LS98}. 

The dynamics of non-perfect conductor threads under the action of radial electric fields have received some attention as well. \citet{CMCP11a} simulated the breakup of a viscous thread covered with ionic surfactants and subject to a radial electric field, taking into account charge relaxation phenomena. They found that the pinching solution tends to the self-similar dynamics of a clean viscous thread at pinching. \citet{LGPH15} showed numerically that electrokinetic effects may considerably alter the distribution of electric charges between primary and satellite droplets, but they do not alter the hydrodynamic balances leading to the asymptotic scaling law of breakup. \citet{W12b} solved the leaky-dielectric model to examine the breakup of a low-conductivity viscous thread immersed in a low-conductivity viscous bath under the action of a radial electric field. The results indicate that the asymptotic behavior is affected by the electric field in this case. The author claimed that the electric shear stress promotes the formation of multiple satellite droplets, analogously to what happens to a surfactant-covered thread under the action of the Marangoni stress. The tangential electrostatic force also plays a key role in the formation of quasi-spike structures during the breakup of viscoelastic electrically charged jets \citep{LKYY19}. \citet{LST16} showed that the local dynamics of a jet in a radial electric field may be altered when the conduction is weak.

The breakup of liquid threads in axial electric fields has been studied on fewer occasions than in the radial configuration. However, the evolution of a leaky-dielectric jet subject to an axial electric field is probably the most interesting configuration due to its relationship with the electrospray cone-jet mode \citep{MG20}. The linear problem exhibits a rich phenomenology depending on the choice of the governing parameters. \citet{S71c} showed that the interaction between the electric and viscous stresses at the electrohydrodynamic boundary layer next to the interface can enhance both asymmetric and axisymmetric instabilities in nearly-inviscid threads. On the contrary, the shear caused by the electric field next to the interface can also suppress capillary instabilities \citep{M94b}. If this shear is sufficiently large, it can excite electrical modes. \citet{M96} showed that all axisymmetric temporal modes can be stabilized for suitable values of the electric field and charges carried by a viscous jet. \citet{CJ08} and \citet{XYQF17} obtained similar results for a viscoelastic jet. \citet{RVHMG20} have recently analyzed the elasto-capillary regime arising during the breakup of an electrified viscoelastic liquid bridge with low conductivity. To the best of our knowledge, the nonlinear breakup of a Newtonian jet/thread immersed in an axial electric field has been studied neither numerically nor experimentally. In this work, we combine high-speed imaging and simulation of the leaky-dielectric model to analyze this problem.

As mentioned above, the driving capillary pressure $\gamma/R_{\textin{min}}$ diverges as the free surface approaches the pinch-off. The resulting force is balanced by inertia, viscosity, or inertia and viscosity depending on the universal regime adopted by the system. The divergence of the hydrodynamic forces at the pinching point may suggest that the forces resulting from an axial electric field will be subdominant sufficiently close to that point. However, the voltage applied to the filament essentially decays along the vanishing axial characteristic length $\ell$ as the free surface approaches the pinch-off. The polarization stress is proportional to $(E^*_t)^2$, where $E^*_t\sim V/\ell$ is the axial (tangential) electric field in the pinching region and $V$ the applied voltage. Therefore, this force also diverges in the pinching point. A natural question is how the competition between the hydrodynamic and electric singularities is resolved. In this paper, we will address this question both numerically and experimentally.

\section{Experimental method}
\label{sec2}

\begin{figure}
\begin{center}
\includegraphics[width=0.9\linewidth]{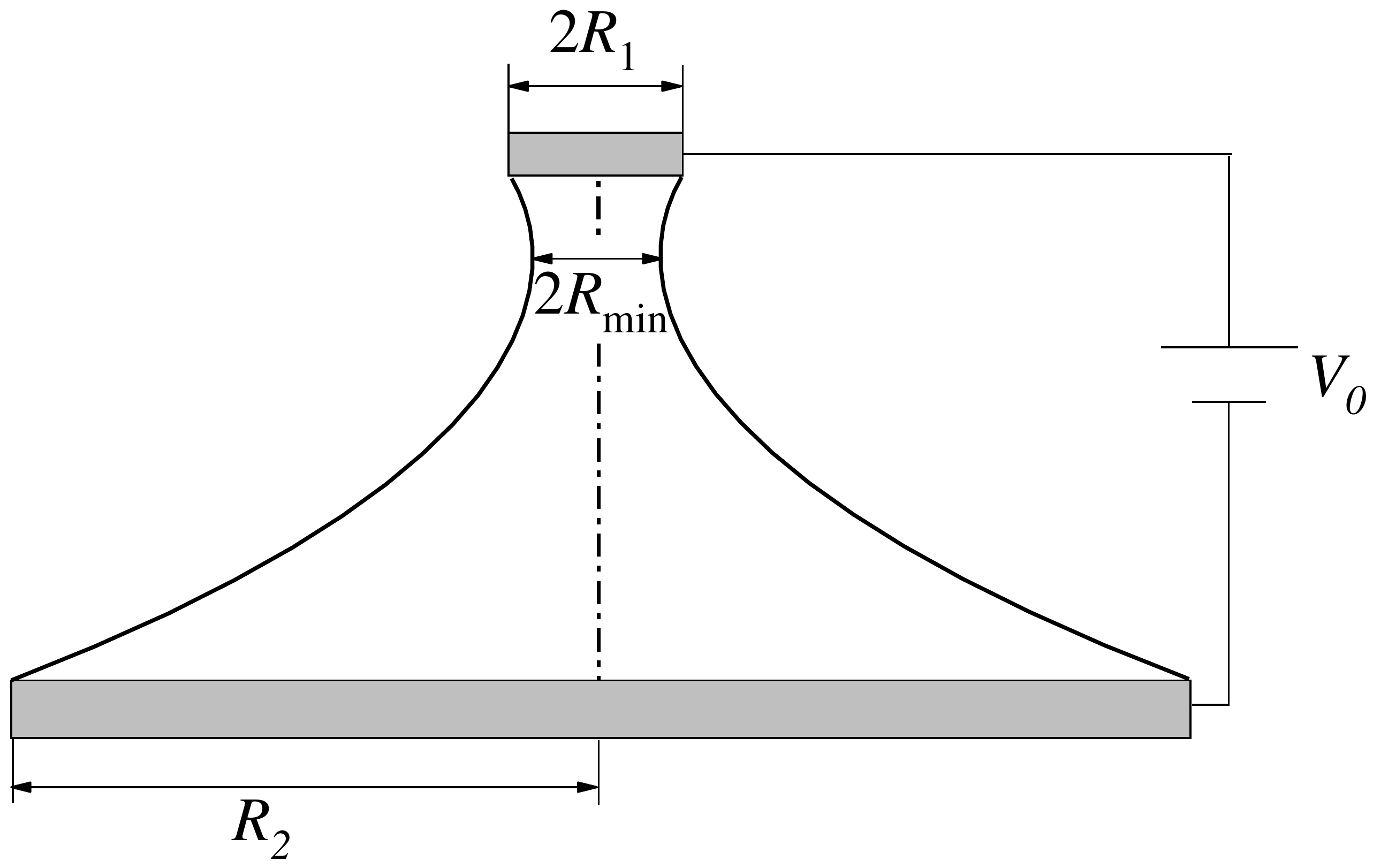}
\end{center}
\caption{Sketch of the experimental configuration}
\label{sketch2}
\end{figure}

To study the effect of the axial electric field on the breakup of a liquid filament, we chose the liquid bridge configuration sketched in Fig.\ \ref{sketch2}. We placed vertically and coaxially a steel capillary and a copper disk of radii $R_1=115$ $\mu$m and $R_2=2$ mm, respectively. These two elements were fixed to high-precision orientation systems to facilitate their correct alignment. A liquid bridge was formed between them by injecting liquid through the upper capillary with a syringe pump (PHD 2000, {\sc Harvard apparatus}). The upper capillary remained still in the course of the experiment, while the lower disk was moved down at a constant speed with a vertical motorized stage (Z825B connected to KDC101, {\sc Thorlabs}). A DC voltage drop $V_0$ was set between the upper capillary and lower (grounded) disk using a function generator ({\sc Keysight}, Model 33210A 10MHz) connected a voltage amplifier ({\sc Trek}, Model 5/80). The submillimeter size of the upper capillary favors the reproducibility of the pinching point location. As will be explained below, the experimental setup included both a high-speed camera and an optical trigger, which prevented us from acquiring images with two perpendicular optical axes. For this reason, the liquid bridge eccentricity could not be checked. Nevertheless, the large value of the lower disk radius made deviations from the axisymmetric configuration irrelevant.

Digital images of the liquid bridge consisting of 924$\times$768 pixels were acquired at $5\times 10^6$ fps using an ultra-high-speed video camera ({\sc Kirana}-5M) equipped with optical lenses (12X {\sc Navitar}) and a microscope objective (10X {\sc Mitutoyo}). The magnification was 250$\times$, which resulted in 0.12 $\mu$m/pixel. The camera could be displaced both horizontally and vertically using a triaxial translation stage with one of its horizontal axes motorized ({\sc Thorlabs} Z825B) and controlled by the computer, which allowed us to set the droplet-to-camera distance with an error smaller than 200 nm. The camera was illuminated with a laser (SI-LUX 640, {\sc Specialised Imaging}) synchronized with the camera, which reduced the effective exposure time down to 100 ns. The camera was triggered by an optical trigger (SI-OT3, {\sc Specialised Imaging}) equipped with optical lenses and illuminated with cold white backlight. The camera triggered the function generator, and, therefore, the first image approximately corresponds to the instant at which the voltage drop was established. All the elements of the experimental setup were mounted on an optical table with a pneumatic anti-vibration isolation system to damp the vibrations coming from the building.

\begin{figure}
\begin{center}
\includegraphics[width=\linewidth]{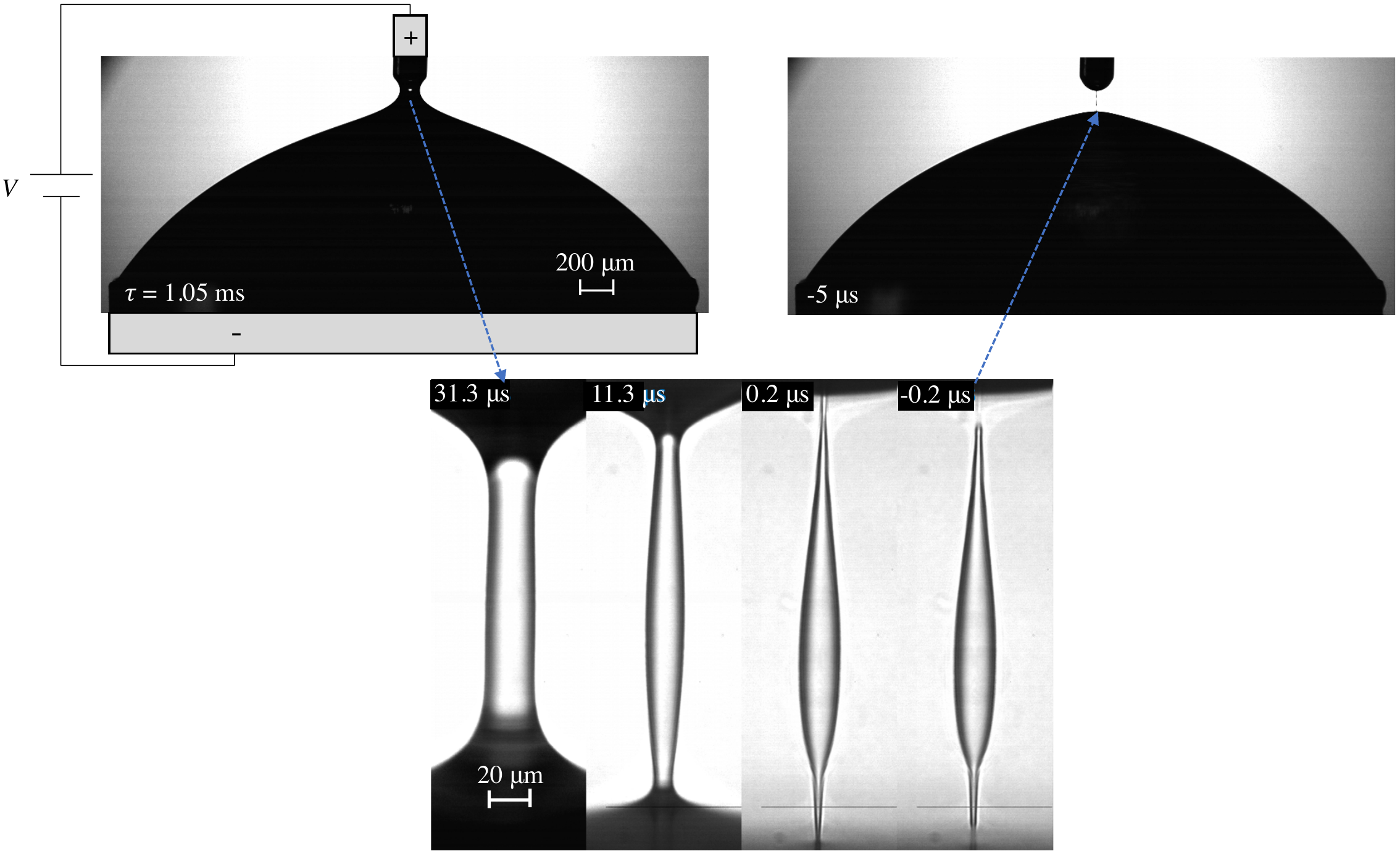}
\end{center}
\caption{Images showing the evolution of a liquid bridge of glycerin-water mixture 30/70\% (w/w) for $V_0=400$ V. The labels indicate the time to the pinching $\tau$.}
\label{evol}
\end{figure}

In the experiments, a liquid bridge of volume 8.3 mm$^3$ was formed between the upper capillary and lower disk, separated initially by a distance around 1.5 mm (Fig.\ \ref{evol}). The upper and lower triple contact lines were pinned to the edges of the capillary and disk, respectively. The liquid bridge was stretched by moving the lower disk away from the upper capillary at the speed 0.1 mm/s. When the maximum length stability limit was reached, the liquid bridge broke up. The breaking process consisted of two phases: (i) the liquid bridge deformation, which gives birth to a thin filament between the upper and lower parent drops, and (ii) the thinning of that filament until the free surface pinches. A DC voltage drop $V_0$ was established between the two electrodes during the last phase. In this way, we avoid the liquid heating due to the Joule effect, and we allow the liquid filament to form. In fact, if the voltage is applied during the first phase mentioned above, the central part of the liquid bridge bulges, and the breakup does not occur. In the experiments, we increased the voltage until the filament could not form. The maximum voltage depends on the liquid properties. Images of the filament were taken during the last phase of the breakup process mentioned above. The images were processed with a sub-pixel resolution technique to determine the free surface position. We calculated the liquid bridge minimum radius $R_{\textin{min}}$ from the free surface contour detected in the images. 

The fluids used in the experiments were mixtures of water and glycerine. We use the notation WXXGYY to refer to a water (W)/glycerine (G) mixture with the proportion XX/YY\% in weight. We used deionized water for synthesis ({\sc Millipore}, {\sc Sigma-Aldrich}) with a conductivity $K=4.3\times 10^{-4}$ S/m at 20 $^{\circ}$C, and glycerol, 99\% for synthesis ({\sc PanReac} {\sc AppliChem}, ITW Reagents). Glycerol of this high purity can be regarded as a quasi-dielectric liquid. In fact, its conductivity is smaller than $10^{-8}$ S/m, the minimum value measurable with our conductometer. We also conducted experiments with octanol, alcohol with lower conductivity and permittivity than those of the water/glycerine mixtures and frequently used in electrospray. 

The properties of the working liquids are shown in Table \ref{tab1}. The density was measured with a pycnometer Glassco of 10 ml, the viscosity was determined with a Fungilab Cannon-Fenske viscometer, the surface tension was measured with the TIFA-AI method \citep{FMC07}, and the electrical conductivity and permittivity were obtained with a parallel-plate cell connected to a LCR meter base \citep{KZ16}. The table also shows the value of the the Ohnesorge number Oh$=\mu(R_1\rho\gamma)^{-1/2}$, the electric relaxation time $t_e=\beta \varepsilon_o/K$ ($\beta$ is the liquid permittivity relative to the vacuum permittivity $\varepsilon_o$), and the electric field scale $E_o=\left(\gamma^2\rho K^2/\varepsilon_o^5\right)^{1/6}$ in electrospray \citep{G97a}.

\begin{table*}[ht]
    \centering
    \begin{tabular}{|c|c|c|c|c|c|c|c|c|}
     \hline
     Liquid & $\rho$ (kg/m$^3$) & $\mu$ (mPa s) & $\gamma$ (mN/m) & Oh & $K$ ($\mu$S/m) & $\beta$ & $t_e$ ($\mu$s) & $E_o$ (MV/m) \\
     \hline
             W70G30 & 1061 & 2.4 & 57 & 0.03 & 311 & 69 & 1.96 & 135 \\
     \hline
             W44G56 & 1133 & 8.3 & 60 & 0.084 & 180 & 69.7 & 3.43 & 118 \\
     \hline
             W30G70 & 1169 & 21.6 & 64 & 0.23 & 57 & 62.1 & 9.7 & 81\\
     \hline
            Octanol & 827 & 7.20 & 23.5 & 0.152 & 2.6 & 10 & 34 & 19.6 \\
      \hline
    \end{tabular}
    \caption{Density $\rho$, viscosity $\mu$, surface tension $\gamma$, Ohnesorge number Oh$=\mu(R_1\rho\gamma)^{-1/2}$, electrical conductivity $K$, relative permittivity $\beta$, electric relaxation time $t_e=\beta \varepsilon_o/K$, and electric field scale in electrospray, $E_o=\left(\gamma^2\rho K^2/\varepsilon_o^5\right)^{1/6}$.}
    \label{tab1}
\end{table*}

The water/glycerine mixtures described above can be regarded as leaky-dielectric (low-conductivity) and ``molecularly simple" liquids. In fact, they exhibit a Newtonian behavior for the times to the pinching analyzed in this work \citep{RPVHM19}, contrary to what occurs to, e.g., silicone oils. In addition, the relatively small size of the water and glycerine molecules renders the permittivity constant for the time scales analyzed in the experiments. As will be seen, the electric relaxation time is sufficiently small for the leaky-dielectric approximation to be valid over the time analyzed in our experiments. In all the cases, the Ohnesorge number is sufficiently large to avoid the free surface overturning at the pinching point, which would prevent us from taking images of the free surface at that point.

\begin{figure}
\begin{center}
\includegraphics[width=0.7\linewidth]{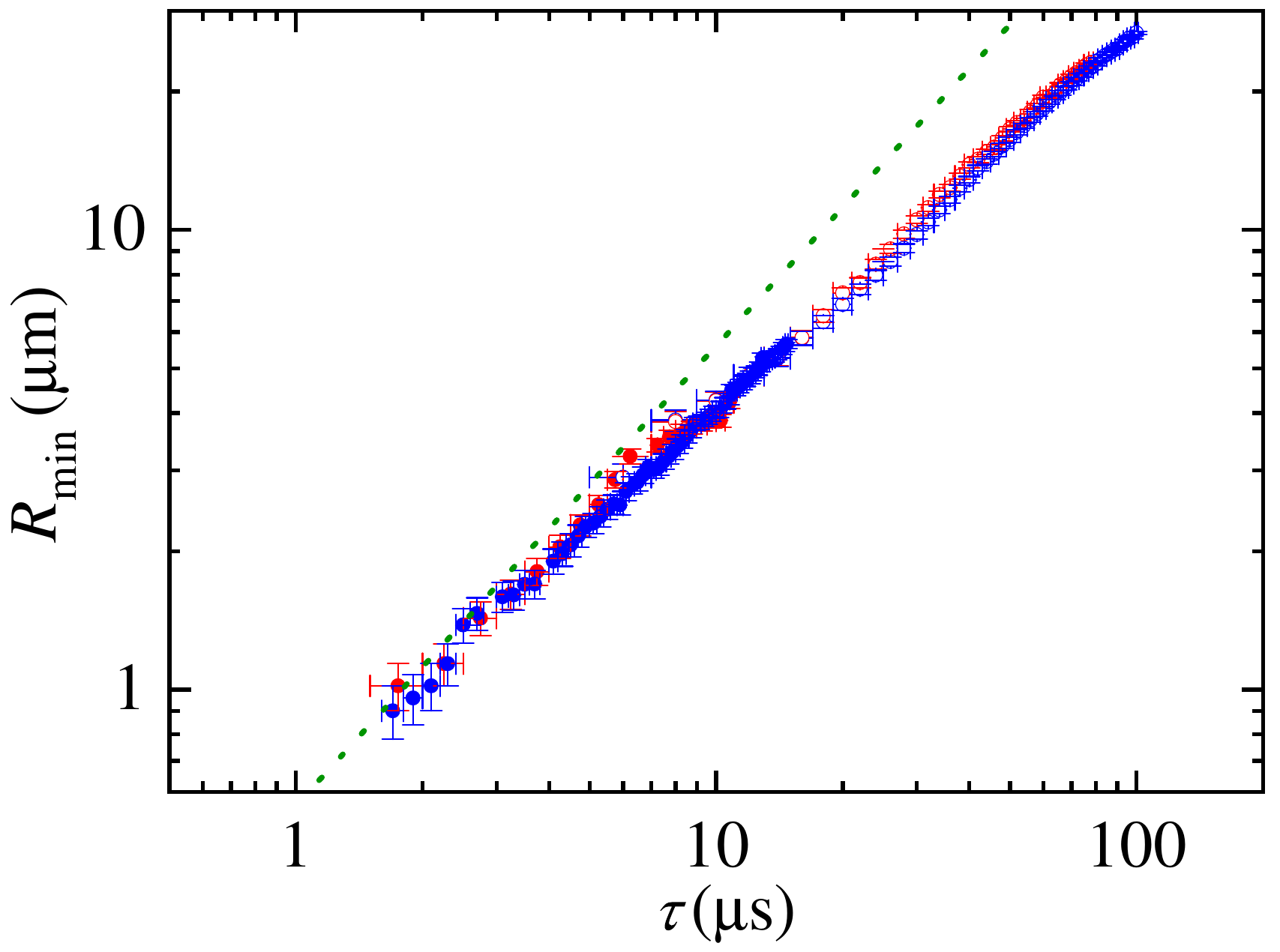}
\end{center}
\caption{$R_{\textin{min}}(\tau)$ for W44G56 and $V_0=0$. The red and blue symbols correspond to the breakup of a liquid bridge and pendant droplet, respectively. The open and solid symbols correspond to experiments conducted with different magnifications. The dotted line is the scaling law $R_{\textin{min}}(\tau)=0.0709\  \gamma/\mu\ \tau$ \citep{P95}.}
\label{Rmin}
\end{figure}

Figure \ref{Rmin} shows $R_{\textin{min}}$ as a function of the time to the pinching, $\tau$, during the breakup of the liquid bridge described in Sec.\ \ref{sec2} and a pendant droplet hanging on the upper capillary (i.e., removing the lower disk). The results correspond to W44G56 in the absence of the electric field. The open and solid symbols correspond to experiments conducted with different magnifications. The agreement between these results shows the high degree of reproducibility of our experiments. The agreement between the radii measured with the liquid bridge and pendant droplet shows the universality of the results for the interval $R_{\textin{min}}(\tau)\lesssim 20$ $\mu$m analyzed in the figure. As can be observed, the presence of a lower disk does not affect the system evolution next to the pinching point. The experimental values of $R_{\textin{min}}(\tau)$ agree with the viscous scaling law (\ref{vis}) \citep{P95} for $R_{\textin{min}}(\tau)\lesssim 5$ $\mu$m. 

\section{The leaky-dielectric model}
\label{sec3}

\begin{figure}
\begin{center}
\includegraphics[width=0.8\linewidth]{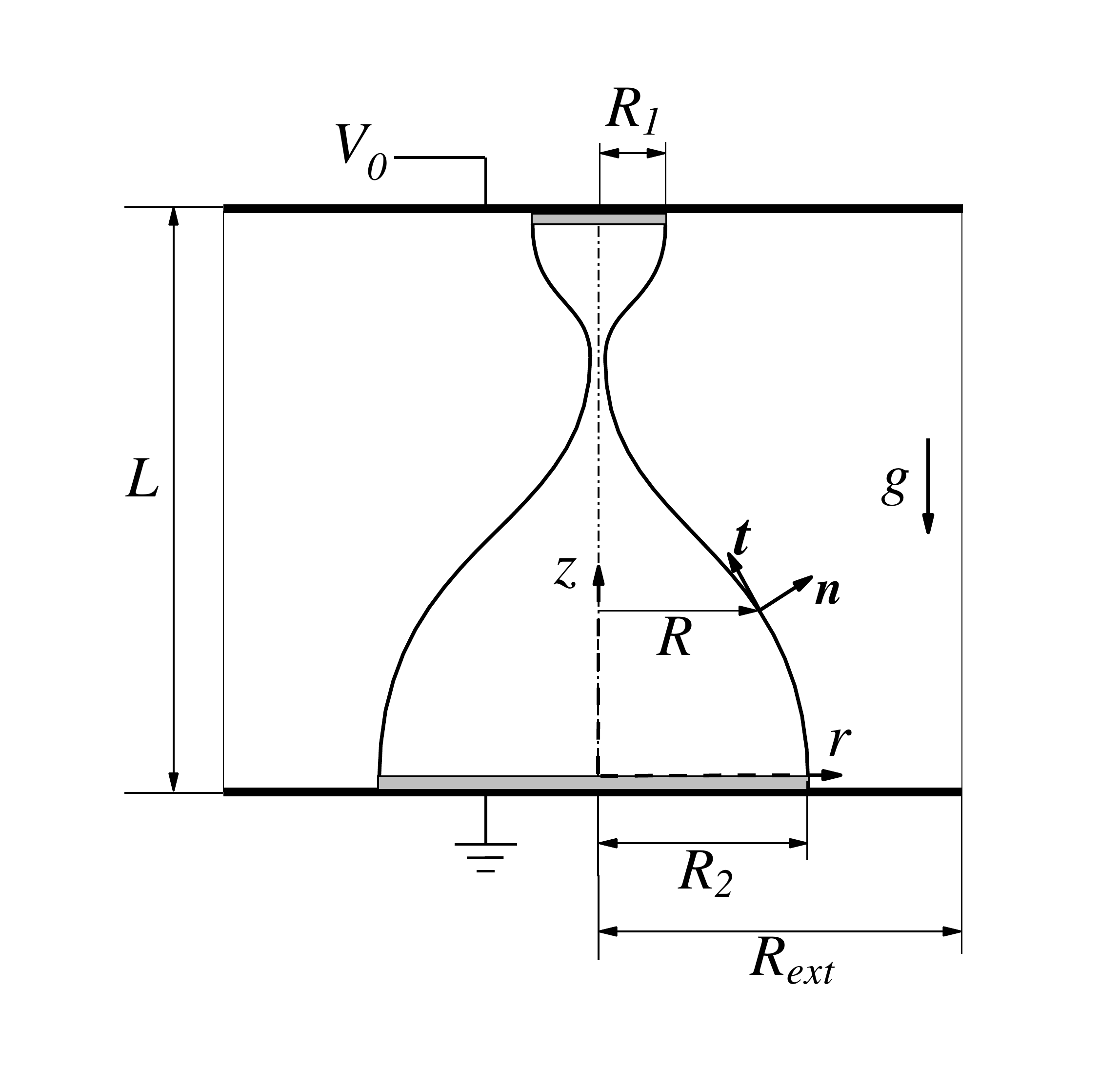}
\end{center}
\caption{Sketch of the numerical problem's formulation}
\label{sketchnum}
\end{figure}

In the theoretical study, we consider an isothermal liquid bridge of constant volume ${\cal V}$, density $\rho$, viscosity $\mu$, electrical conductivity $K$, and relative permittivity $\beta$. The liquid bridge is surrounded by a dielectric fluid of negligible density and viscosity. The electrical permittivity of this fluid is $\varepsilon_o$. The liquid bridge is held vertically by the constant surface tension $\gamma$ between two horizontal electrodes of radius $R_{\textin{ext}}$ and separated by a distance $L$  (Fig.\ \ref{sketchnum}). A constant voltage $V_0$ is applied between the two electrodes. The upper and lower triple contact lines are pinned to the solid surfaces at the distances $R_1$ and $R_2$ ($R_1, R_2\ll R_{\textin{ext}}$) from the liquid bridge axis, respectively.

In this section, all the quantities are made dimensionless with the upper triple contact line radius $R_1$, the liquid density $\rho$, the surface tension $\gamma$, and the voltage drop $V_0$. This choice yields the characteristic scales $t_c=(\rho R_1^3/\gamma)^{1/2}$, $v_c =R_1/t_c$, $p_c=\gamma/R_1$ and $E_c=V_0/R_1$, respectively. For the sake of simplicity, in the rest of this section the symbols represent dimensionless quantities.

The velocity ${\bf v}(r,z,t)=u(r,z,t){\bf e_r}+w(r,z,t){\bf e_z}$ and modified pressure (the hydrostatic pressure plus gravitational potential per unit volume) $p(r,z,t)$ fields are calculated from the continuity and momentum equations
\begin{equation}
\label{e1}
{\boldsymbol \nabla}\cdot {\bf v}=0, 
\end{equation}
\begin{equation}
\label{m}
\frac{\partial {\bf v}}{\partial t}+{\bf v}\cdot {\boldsymbol \nabla}{\bf v}=-{\boldsymbol \nabla}p+{\boldsymbol \nabla}\cdot {\bf T},
\end{equation}
where ${\bf T}=\text{Oh}[{\boldsymbol \nabla}{\bf v}+({\boldsymbol \nabla}{\bf v})^T]$ is the viscous stress tensor. 

In the leaky-dielectric model \citep{MT69,S97,SY15}, the bulk net free charge is assumed to be negligible, and, therefore, the electric potentials $\phi^i$ and $\phi^o$ in the inner and outer domains obey the Laplace equation 
\begin{equation}
{\boldsymbol \nabla}^2\phi^{i,o}=0\label{el1}.
\end{equation}
The inner and outer electric fields ${\bf E}^{i,o}= E^{i,o}_r {\bf e_r}+ E^{i,o}_z {\bf e_z}$ are calculated as ${\bf E}^{i,o}={\boldsymbol \nabla} \phi ^{i,o}$.

The free surface location is defined by the equation $r=R(z,t)$. The boundary conditions at that surface are:
\begin{equation}
\frac{\partial R}{\partial t}+R_z w-u=0,
\label{int1}
\end{equation}
\begin{eqnarray}
&&-p+Bz-\frac{RR_{zz}-1-R_z^{2}}{R(1+R_z^{2})^{3/2}}+{\bf n}\cdot {\bf T}\cdot {\bf n}=\nonumber\\
&&\frac{\chi}{2}\left[(E_n^o)^2-\beta (E_n^i)^2\right]+\chi\frac{\beta-1}{2}(E_t)^2,
\label{int3}
\end{eqnarray}
\begin{equation}
{\bf t}\cdot {\bf T}\cdot {\bf n}=\sigma E_t, \label{int2}
\end{equation}
where $R_z\equiv dR/dz$ and $R_{zz}=d^2R/dz^2$, $B=\rho g R_1^2/\gamma$ is the gravitational Bond number, $g$ the gravitational acceleration, ${\bf n}$ is the unit outward normal vector, $\chi=\varepsilon_o V_0^2/(R_1\gamma)$ is the electric Bond number, ${\bf t}$ is the unit vector tangential to the free surface meridians, and $\sigma$ is the surface charge density. Equation (\ref{int1}) is the kinematic compatibility condition, while Eqs.\ (\ref{int3}) and (\ref{int2}) express the balance of normal and tangential stresses on the two sides of the free surface, respectively. The right-hand sides of these equations are the Maxwell stresses resulting from both the accumulation of free electric charges at the interface and the jump of permittivity across that surface. The pressure in the outer medium has been set to zero. 

The electric field at the free surface and the surface charge density are calculated as
\begin{equation}
E_n^i=\frac{-R_z E^i_{z}+ E^i_r}{\sqrt{1+R_z^2}}, \quad E_n^o=\frac{-R_zE^o_z+ E^o_r}{\sqrt{1+R_z^2}},
\label{int4}
\end{equation}
\begin{equation}
E_t= \frac{R_z E^o_r+E^o_z}{\sqrt{1+R_z^2}}=\frac{R_zE^i_r+E^i_z}{\sqrt{1+R_z^2}}, 
\label{nose}
\end{equation}
\begin{equation}
\sigma=\chi(E_n^o-\beta E_n^i).
\label{int6}
\end{equation}
It must be noted that the continuity of the electric potential across the free surface, $\phi^i=\phi^o$, has been considered in Eq.\ (\ref{nose}).

The free surface equations are completed by imposing the surface charge conservation at $r=R(z,t)$,
\begin{equation}
\frac{\partial \sigma}{\partial t}+\sigma v_n({\boldsymbol \nabla}\cdot {\bf n})+\boldsymbol{\nabla_s}\cdot (\sigma{\bf v_s})=\chi \alpha E_n^i,
\label{int7}
\end{equation}
where $\boldsymbol{\nabla_s}$ is the tangential intrinsic gradient along the free surface, ${\bf v_s}=v_t {\bf t}$ is the projection of the velocity of a free surface element onto the free surface, and $\alpha=K\left[\rho R_1^3/(\gamma \varepsilon_o^2)\right]^{1/2}$ is the dimensionless electrical conductivity. The diffusion term has been neglected because it is usually much smaller than the other terms \citep{GLHRM18}.

The anchorage conditions $R=1$ and $R=R_2/R_1$ are set at $z=\Lambda$ and $z=0$, respectively, where $\Lambda=L/R_1$ is the liquid bridge slenderness. The nonslip boundary condition is imposed at the solid surfaces in contact with the liquid. The nondimensional volume $\hat{{\cal V}}={\cal V}/R_1^3$ of the initial configuration is prescribed (and conserved), namely,
\begin{equation}
\pi\int_{0}^{\Lambda} R^2\ dz=\hat{{\cal V}}.\label{volume}
\end{equation}
The surface charge conservation equation (\ref{int7}) is integrated by assuming zero surface charge flux at the triple contact lines. The regularity conditions $E^i_ r = u=w_r =0$ are prescribed on the symmetry axis. We fix the electric potential $\phi^{i,o}=0$ and $\phi_0$ at the lower and upper electrodes, respectively. The linear relationship $\phi^{o}=\phi_0\, z/\Lambda$ is set at the cylindrical lateral surface $r=R_{\textin{ext}}/R_1$.

We start the simulation from a non-electrified liquid bridge at equilibrium with slenderness just below the critical one. We trigger the breakup process at the initial instant by applying a very small gravitational force (i.e., by slightly changing the Bond number value). As also done in the experiments, the voltage drop is applied at some instant before reaching the time interval of interest. Then, we simulate the liquid bridge breakup under the action of the electric field. 

The leaky-dielectric model was solved with a variation of the method described by \citet{HM16a}. The physical domains occupied by the liquid and the outer dielectric medium were mapped onto two rectangular domains through a coordinate transformation. Each variable and its spatial and temporal derivatives appearing in the transformed equations were written as a single symbolic vector. Then, we used a symbolic toolbox to calculate the analytical Jacobians of all the equations with respect to the symbolic vector. Using these analytical Jacobians, we generated functions that could be evaluated in the iterations at each point of the discretized numerical domains. 

The transformed spatial domains were discretized using $n_\eta^{(i)}=25$ and $n_\eta^{(o)}=21$ Chebyshev spectral collocation points \citep{KMA89} in the transformed radial direction $\eta$ of the inner and outer domains, respectively. We applied a stretching function to concentrate points near the interface in the outer domain.  We used $n_\xi=7001$ equally spaced collocation points in the transformed axial direction $\xi$. The axial direction was discretized using fourth-order finite differences. Second-order backward finite differences were used to discretize the time domain. We used an automatic variable time step based on the norm of the difference between the solution calculated with a first-order approximation and that obtained from the second-order procedure. The nonlinear system of discretized equations was solved at each time step using the Newton method. The method is fully implicit.

\begin{figure}
\begin{center}
\includegraphics[width=1\linewidth]{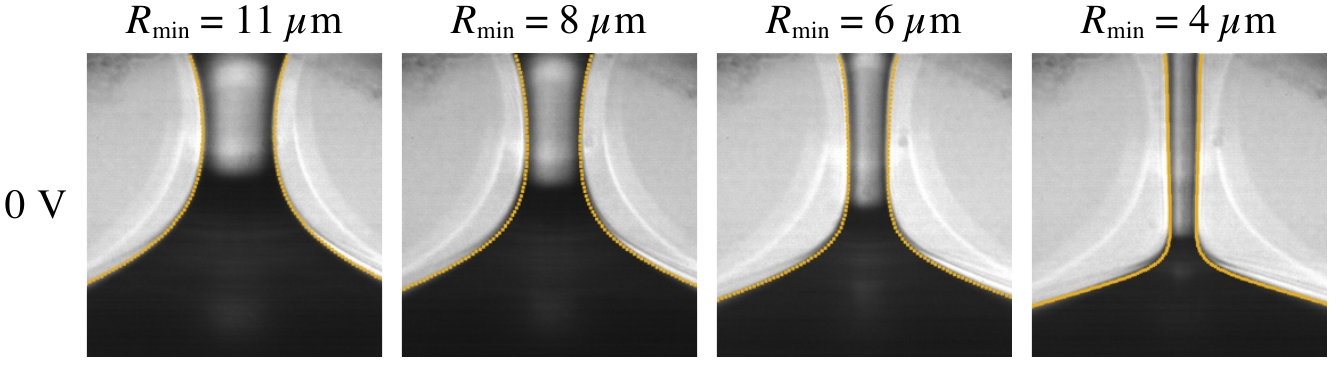}
\end{center}
\caption{Images of W44G56 for $V_0=0$ V. The solid lines are the numerical contours calculated at the instants when $R_{\textin{min}}(t)$ took the same value as that of the experiment.}
\label{vali}
\end{figure}

The comparison between the experimental and numerical filament shapes next to the pinching point (Fig.\ \ref{vali}) shows the accuracy of the numerical simulation in the absence of the electric field. This validation will be completed in Sec.\ \ref{sec42} for electrified filaments. It must be pointed out that our numerical method does not allow us to go beyond the free surface pinch-off. That phase of the system dynamics will be studied experimentally.

\section{Results}
\label{sec4}

\subsection{Experimental results}
\label{sec41}

\begin{figure}
\begin{center}
\includegraphics[width=1\linewidth]{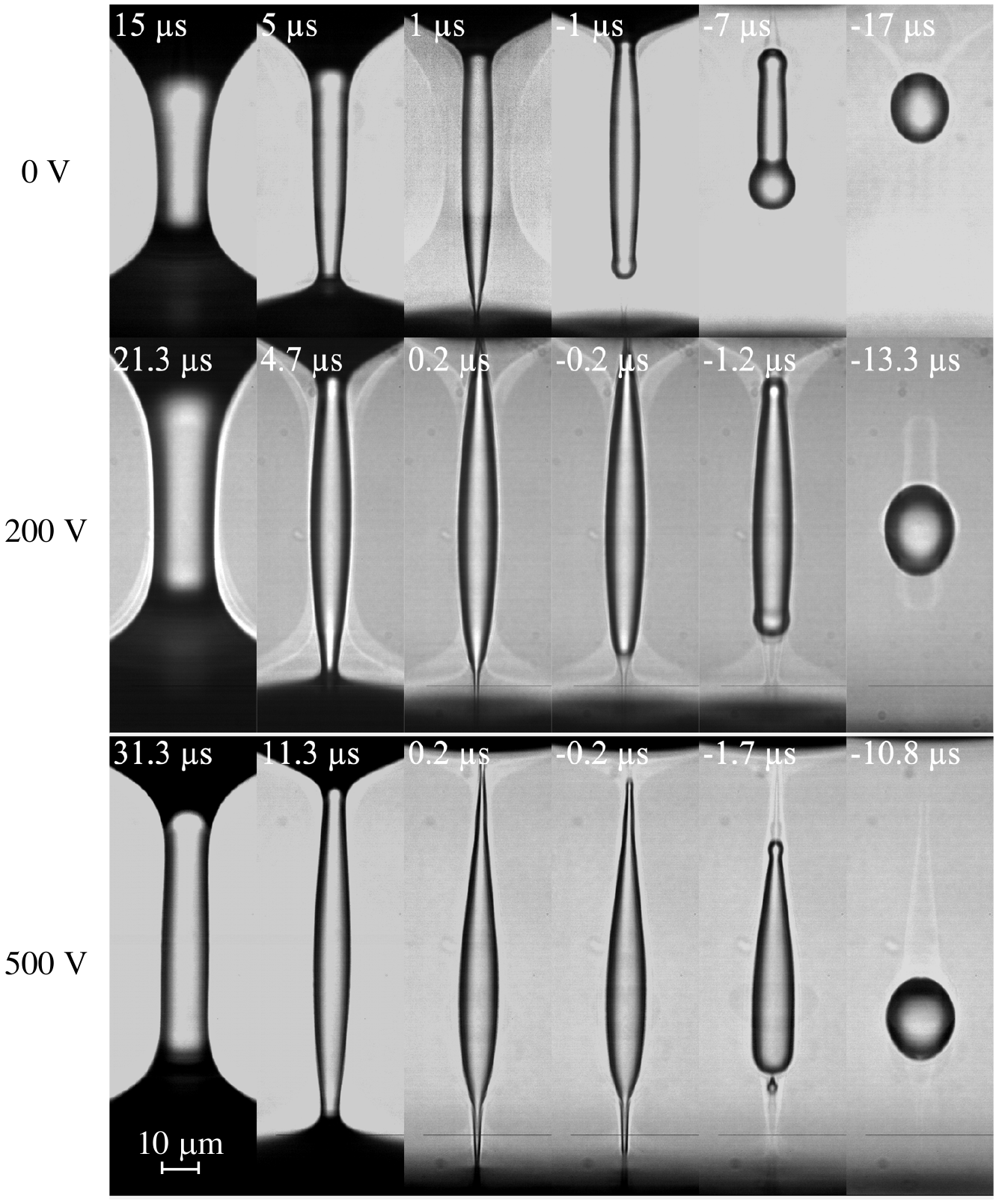}
\end{center}
\caption{Images showing the evolution of a liquid bridge of W70G30 for $V_0=0$, 200, and 400 V. The labels indicate the time to the pinching $\tau$. The voltages $V_0=200$ and 400 V were applied at $\tau=52.8$ and 47.3 $\mu$s, respectively.}
\label{evol1}
\end{figure}

\begin{figure}
\begin{center}
\includegraphics[width=1\linewidth]{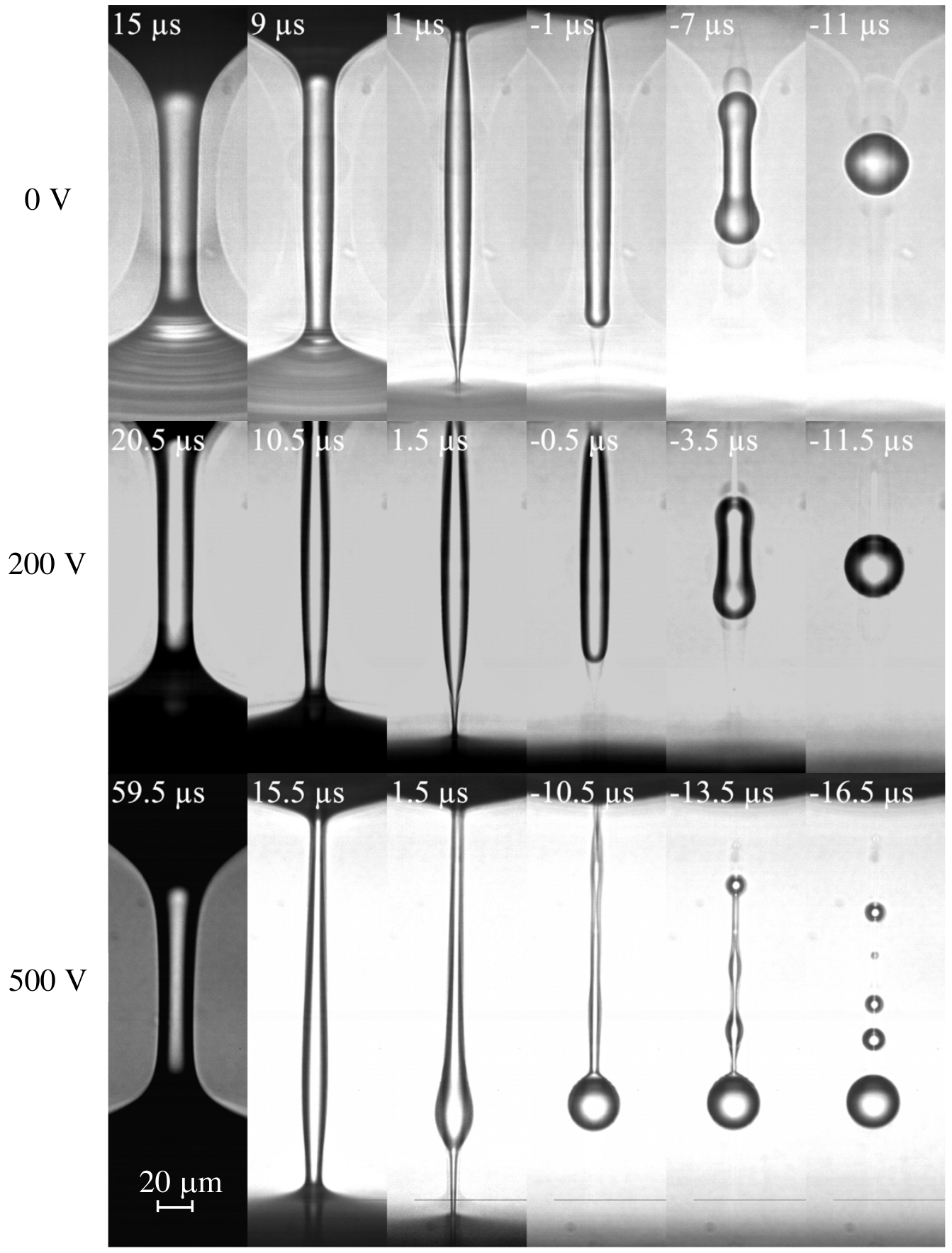}
\end{center}
\caption{Images showing the evolution of a liquid bridge of W44G56 for $V_0=0$, 200, and 500 V. The labels indicate the time to the pinching $\tau$. The voltages $V_0=200$ and 500 V were applied at $\tau=53.5$ and 61.5 $\mu$s, respectively.}
\label{evol2}
\end{figure}

\begin{figure}
\begin{center}
\includegraphics[width=0.85\linewidth]{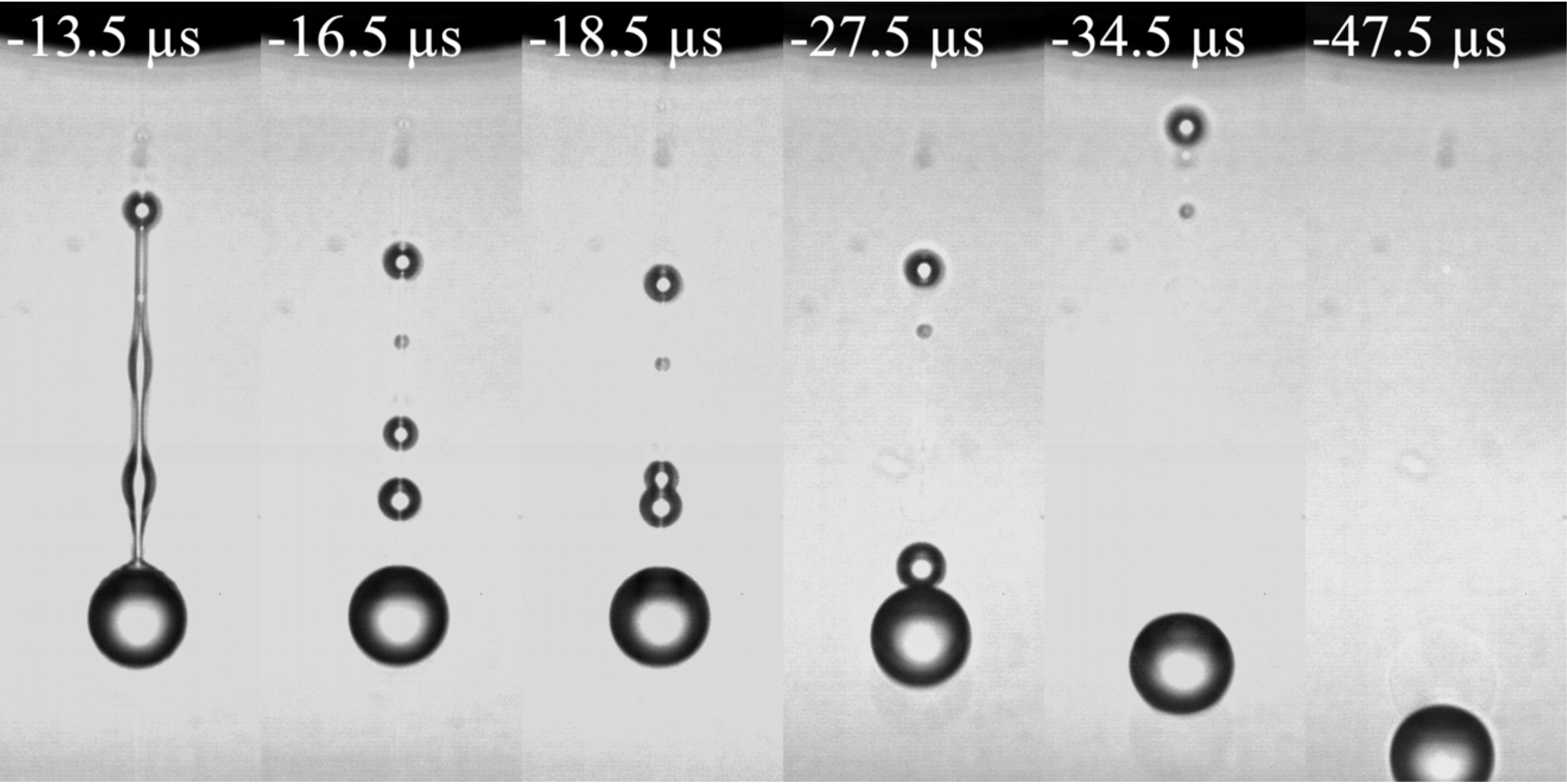}
\end{center}
\caption{Images showing the paths of the droplets created after the pinching of W44G56 for $V_0=500$ V. The labels indicate the time to the pinching $\tau$. The voltage was applied at $\tau=61.5$ $\mu$s.}
\label{tray}
\end{figure}

\begin{figure}
\begin{center}
\includegraphics[width=1\linewidth]{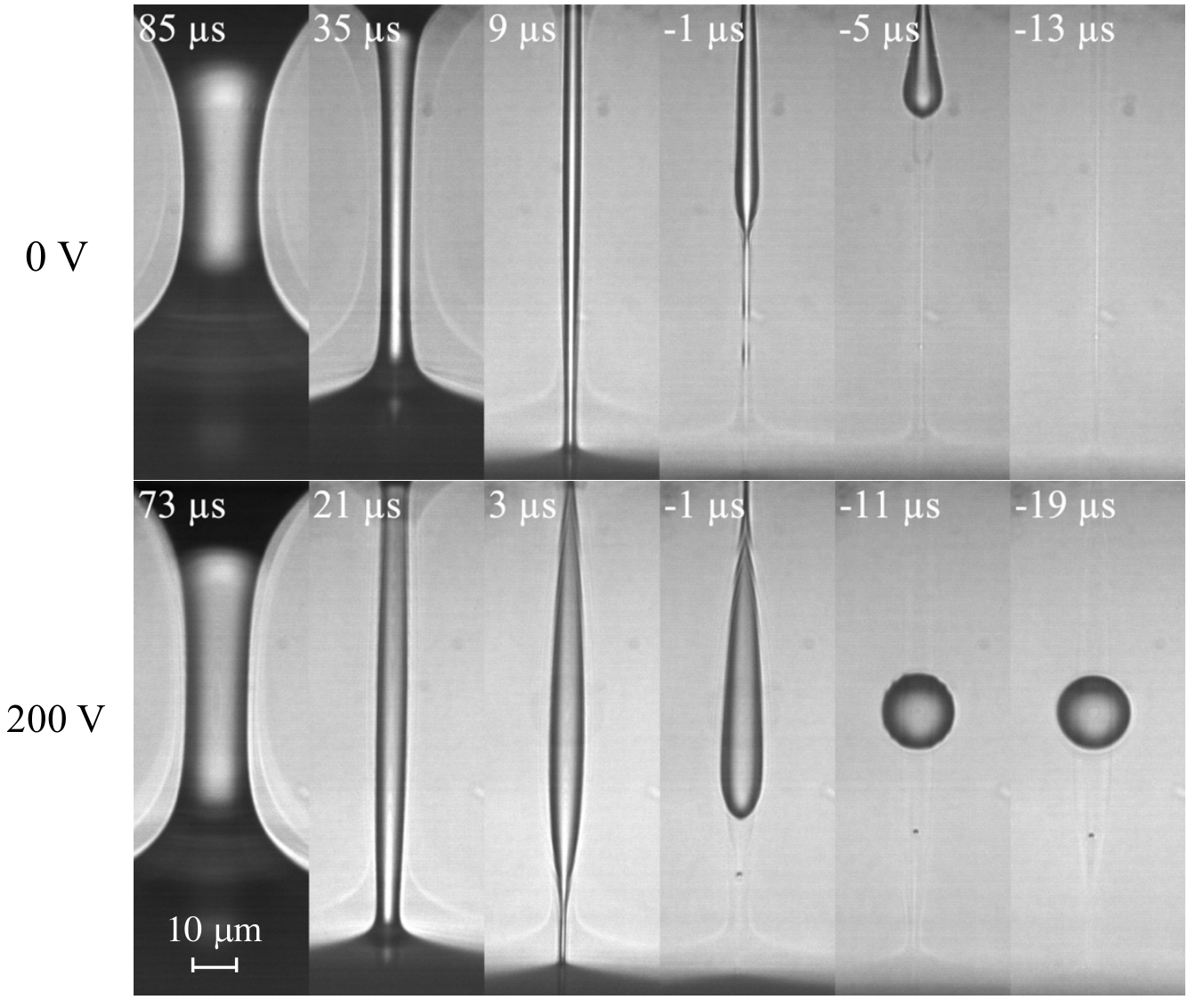}
\end{center}
\caption{Images showing the evolution of a liquid bridge of W30G70 for $V_0=0$ and 200 V. The labels indicate the time to the pinching $\tau$. The voltage $V_0=200$ was applied at $\tau=125$ $\mu$s.}
\label{evol3}
\end{figure}

\begin{figure}
\begin{center}
\includegraphics[width=1\linewidth]{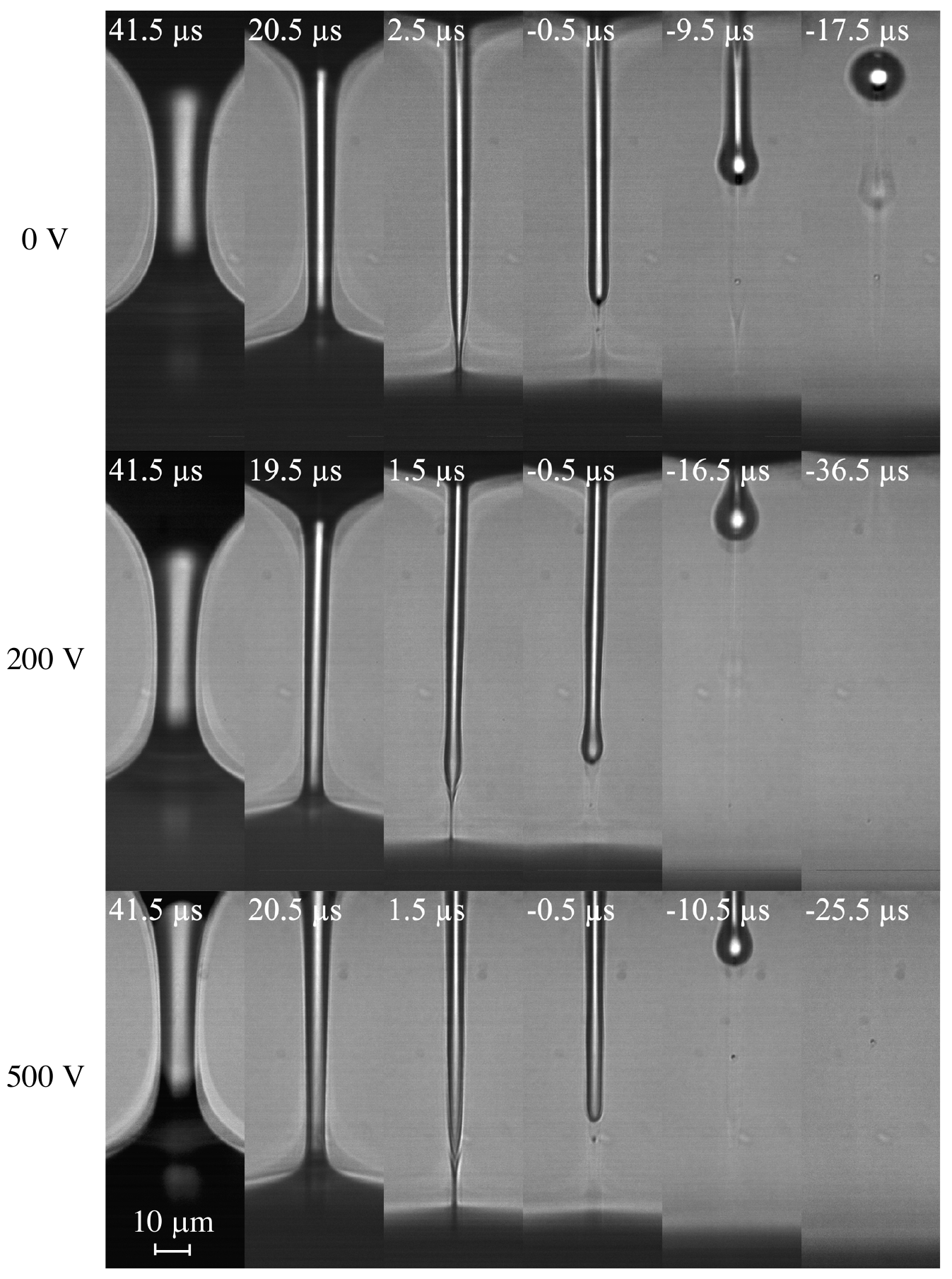}
\end{center}
\caption{Images showing the evolution of a liquid bridge of octanol for $V_0=0$, 200, and 500 V. The labels indicate the time to the pinching $\tau$. The voltages $V_0=200$ and 500 V were applied at $\tau=76.5$ and 64.5 $\mu$s, respectively.}
\label{evol4}
\end{figure}

We examine the qualitative influence of the applied electric field on the liquid bridge breakup in Figs.\ \ref{evol1}-\ref{evol4}. For W70G30 (Fig.\ \ref{evol1}), the electric field significantly affects the size of the satellite droplet formed between the upper and lower parent drops. The dependence of the satellite droplet diameter on the electric field is non-monotonous. Specifically, the diameter increases when the voltage $V_0=200$ V is applied and decreases when the voltage is increased from 200 V up to 400 V. The first pinching takes place at the lower end of the liquid filament for both $V_0=0$ and 200 V, while for $V_0=400$ V the free surface pinches almost simultaneously at the two ends. As will be seen in more detail below, this is attributed to the delay of the pinching process caused by the polarization force, which sharply increases as the free surface approaches the pinch-off. A tiny subsatellite droplet forms during the last stage of the pinching for $V_0=400$ V. 

The difference between the breakup in the absence and presence of the electric field becomes noticeable when a high voltage $V_0=500$ V is applied to W44G56, the liquid with intermediate values of viscosity and electrical conductivity (Fig.\ \ref{evol2}). Due to the action of the electric field, the liquid filament bulges above the lower parent drop. When the filament detaches from the lower drop, that bulge gives rise to a satellite droplet connected to the upper parent drop through a thin filament. Several tiny droplets are produced from the breakup of the retracting filament due to the capillary instability. This process takes place during a time interval of around 5 $\mu$s, while the electric relaxation time in this case is $t_e=3.46$ $\mu$s (see Table \ref{tab1}). Therefore, one expects the electric charge trapped in the filament to distribute over the filament according to the charge polarity. Specifically, the negative/positive charge moves up/down along the fluid thread. The droplets resulting from the filament breakup are non-neutral. The three lower droplets have positive charges and are attracted by the bottom parent drop. The opposite occurs with the two upper ones. The images in Fig.\ \ref{tray} show the electro-coalescence between the three lower droplets. This phenomenon does not occur to the two upper droplets due to the drag force experienced by the smallest one.

Figure \ref{evol3} shows the noticeable effect produced by the electric field on the breakup of a viscous liquid bridge of W30G70. In the absence of the electric field, a quasi-cylindrical filament forms between the two parent drops. A long microfilament forms above the lower parent drop over the time interval $0<\tau<9$ $\mu$s. This phenomenon resembles that observed by \citet{K96}, who described the formation of long microthreads during the breakup of jets around 50 cSt in viscosity. Those microthreads stretched until their diameters fell down below approximately 1 $\mu$m. Then, the thinning process practically stopped, and the microthread broke up due to the capillary instability, giving rise to submicrometer subsatellite droplets. In our experiment without an electric field, the filament detaches from the lower parent drop and completely retracts towards the upper drop. Consequently, this process only releases the subsatellite droplets mentioned above. On the contrary, the electric field makes the liquid accumulate in the central part of the filament. The free surface pinches almost simultaneously at the two filament ends, and a single satellite droplet forms for $V_0=200$ V.

The experiments with octanol show the opposite effect to that observed with W3070. The free surface inevitably pinches at the lower end of the filament in the first place, and the filament retracts towards the upper parent drop before releasing the satellite droplet. The difference between the water-glycerine mixtures and octanol behaviors may be attributed to the lower permittivity of the latter. As will be explained in the numerical section, the polarization force is responsible for delaying the first pinching. This force is not sufficiently intense for octanol owing to its lower permittivity. As can be observed in the images, a submicrometer satellite droplet is produced by the breakup of the microthread formed above the lower parent drop, both in the absence and under the action of the electric field. 

The images displayed in Figs.\ \ref{evol1}-\ref{evol4} show the rich phenomenology arising when a sufficiently intense axial electric field is applied on a liquid filament connecting two equipotential parent drops. We have verified that the effects described above are reproducible, i.e., they always appear when the experiments are repeated under the same experimental conditions. The experimental images show that the last stage before the free surface pinching is ``contaminated" by the formation of subsatellite droplets and other effects \citep{RPVHM19}. These effects affect the breakup time and hinder the search of scaling laws for $R_{\textin{min}}(\tau)$. 

\begin{figure*}
\begin{center}
\includegraphics[width=0.35\linewidth]{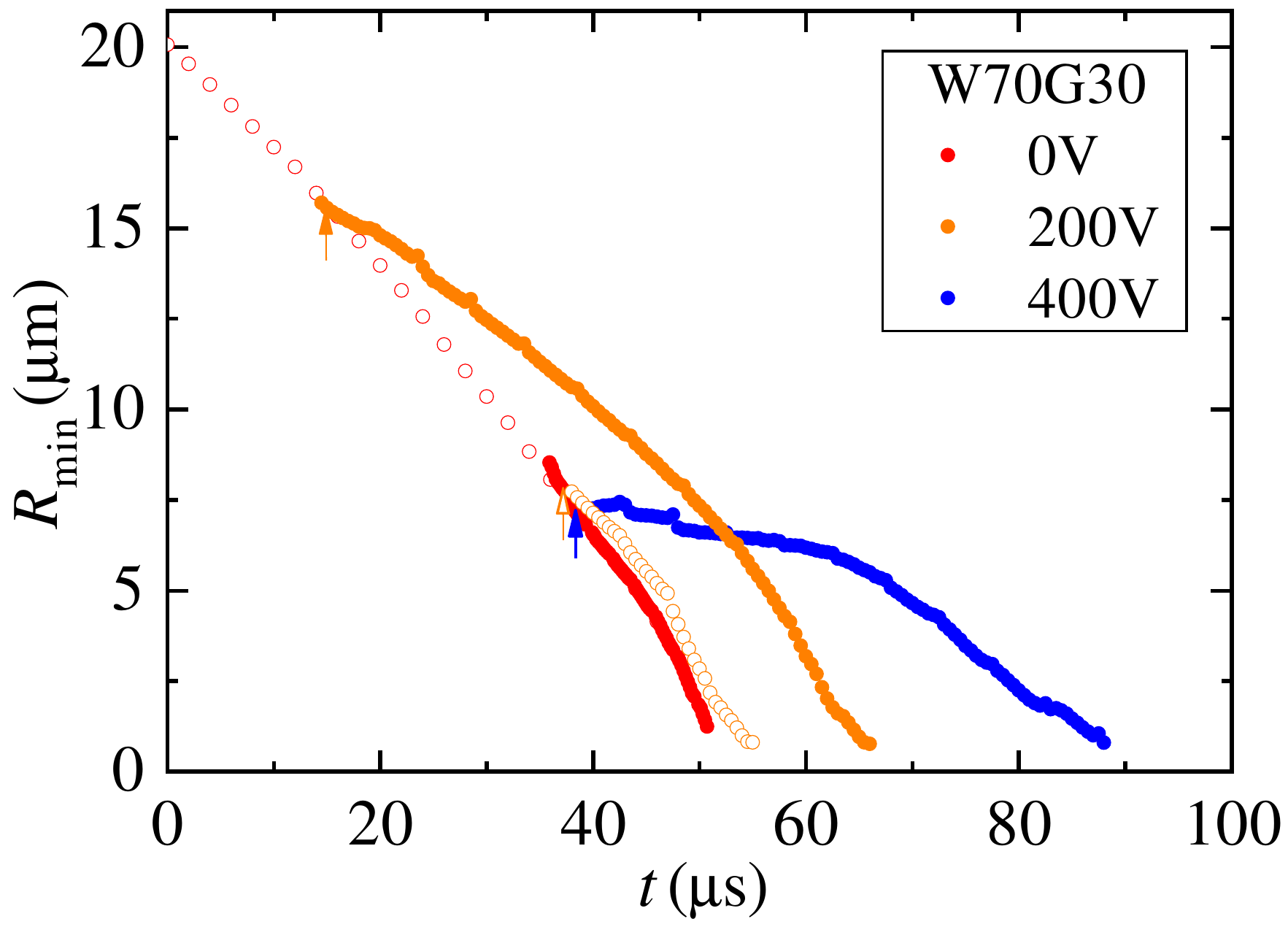}
\includegraphics[width=0.35\linewidth]{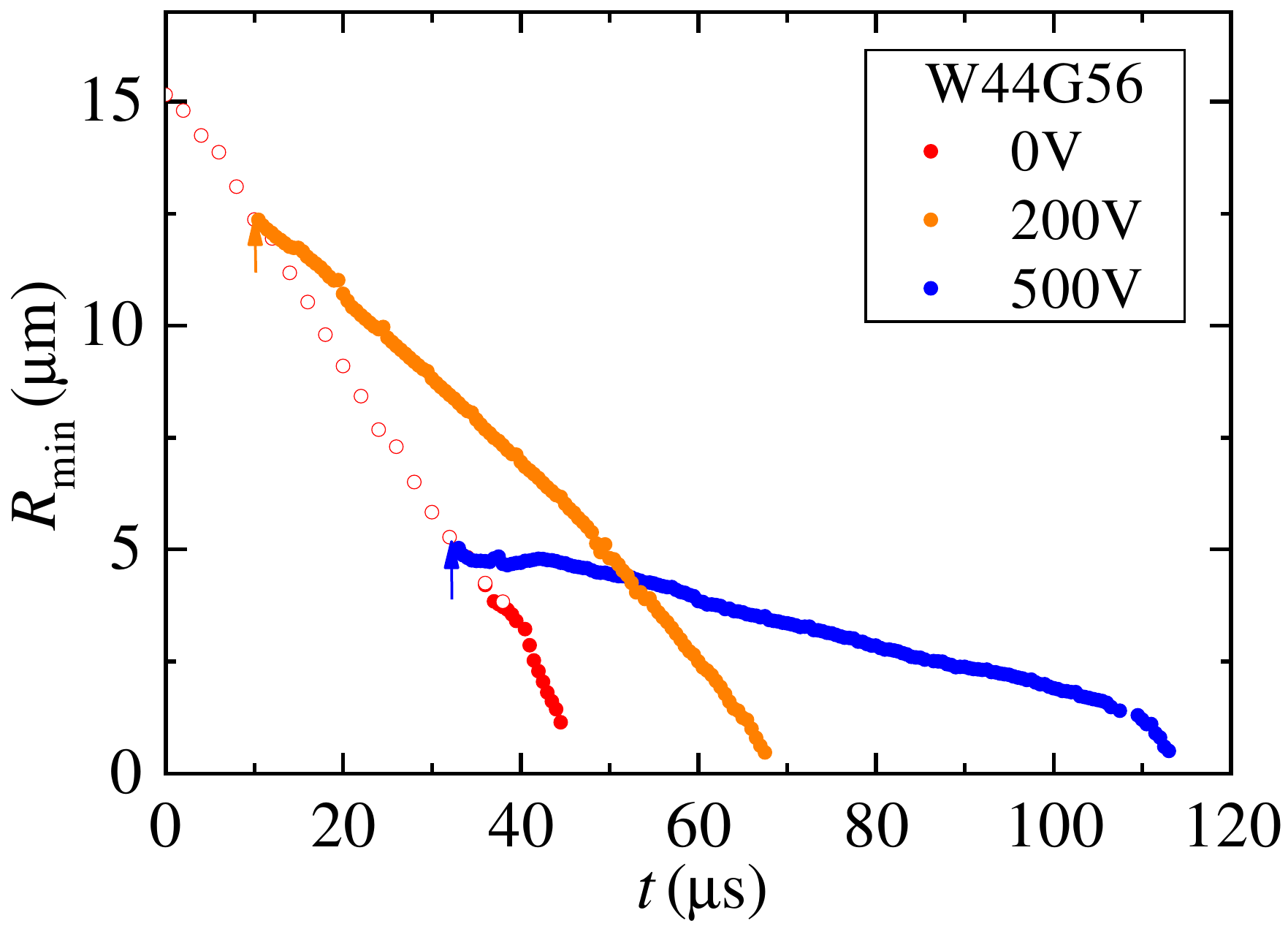}\\
\includegraphics[width=0.35\linewidth]{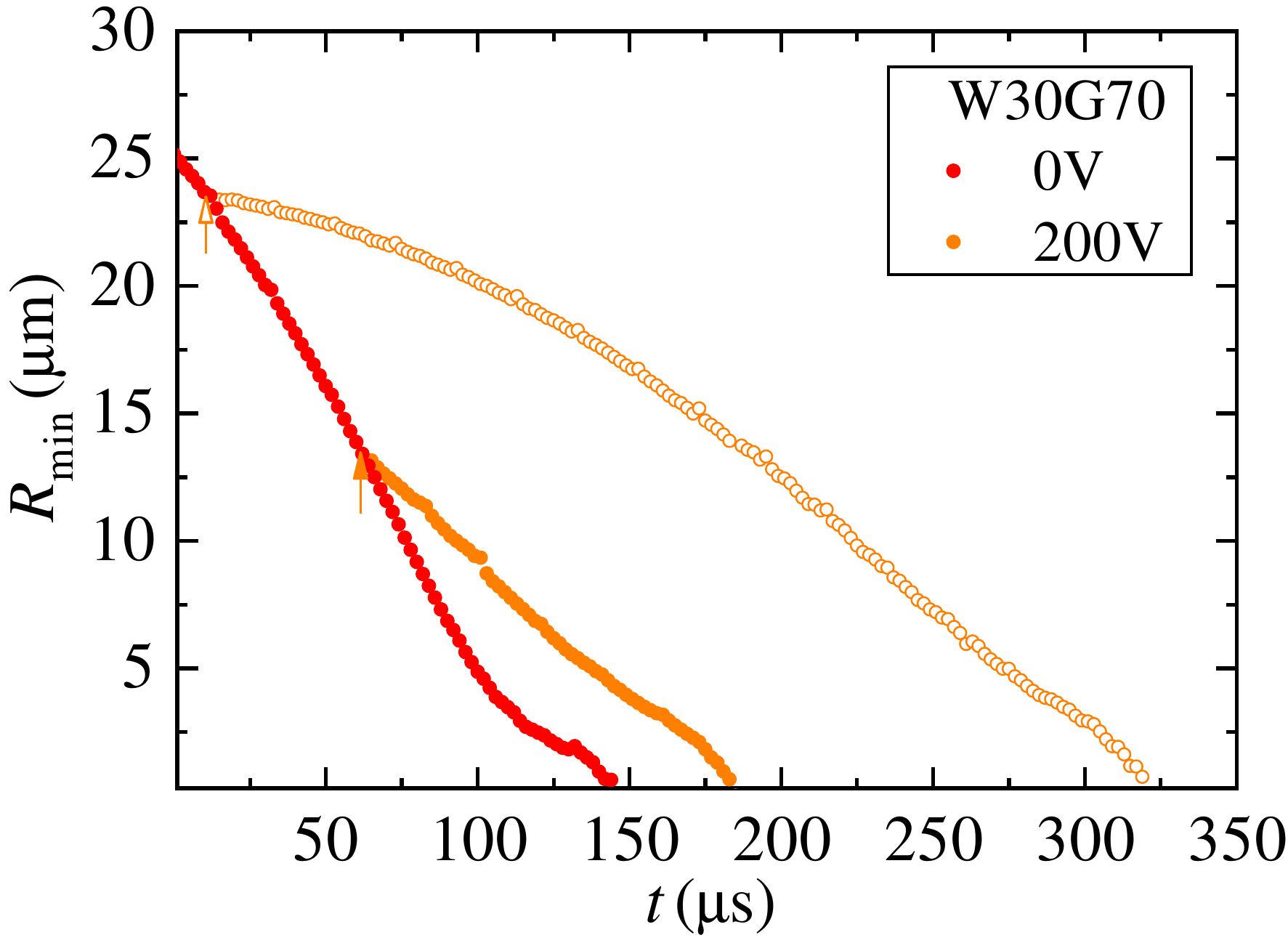}
\includegraphics[width=0.35\linewidth]{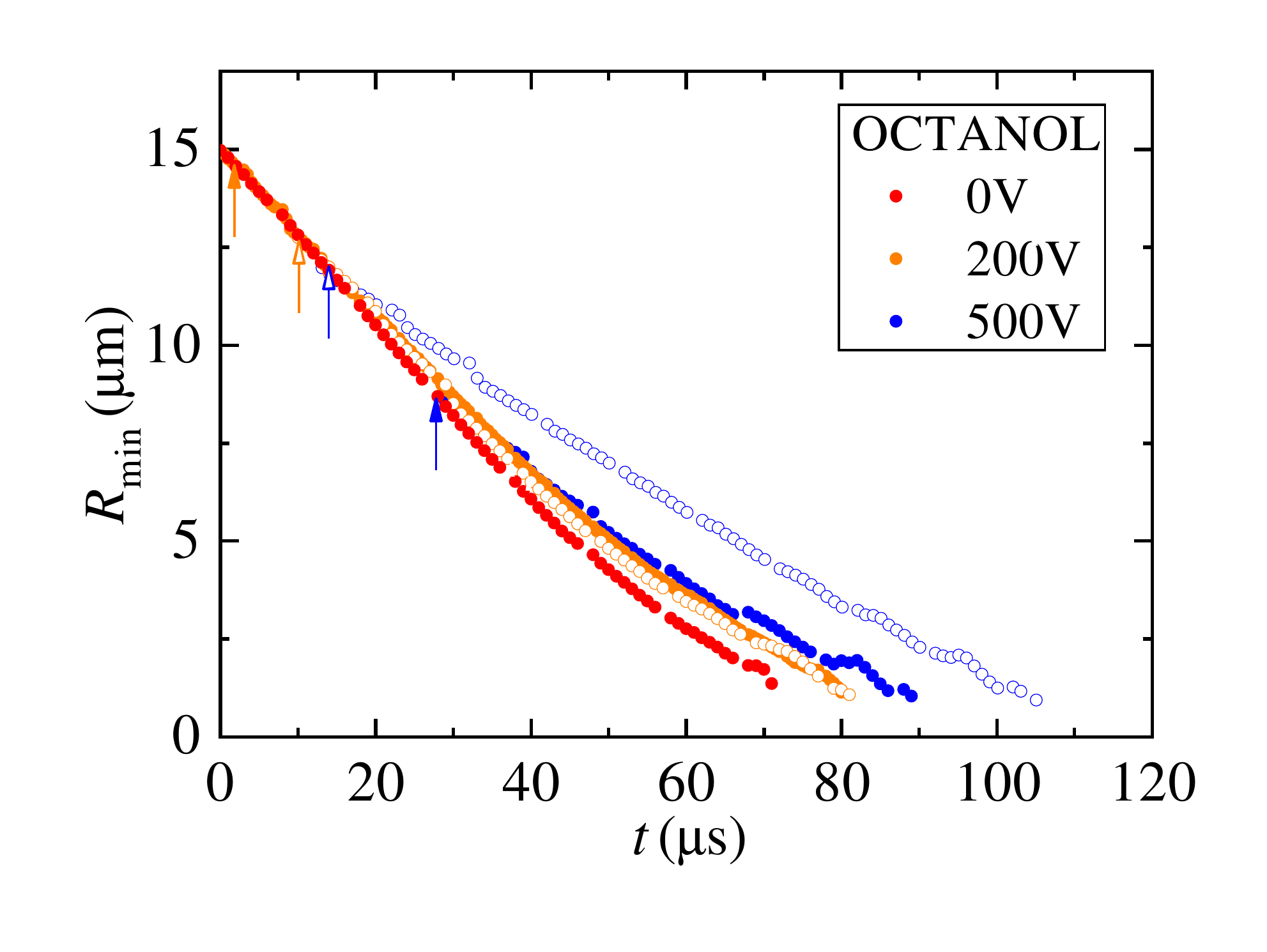}
\end{center}
\caption{$R_{\textin{min}}(t)$ for the four liquids analyzed. The arrows indicate the instants at which the voltage was applied. The arrows with open and solid arrowheads correspond to the open and solid symbols, respectively.}
\label{Rmin0}
\end{figure*}

To conduct a quantitative analysis, we measured the minimum radius of the free surface in the vicinity of the lower parent drop as a function of time. Figure \ref{Rmin0} shows $R_{\textin{min}}(t)$ both in the absence of the electric field and when the voltage drop is established. The arrows indicate the time at which the electric field was applied. The instant $t=0$ has no physical meaning. It corresponds to the instant at which we started taking images in the experiment without the electric field. As mentioned in Sec.\ \ref{sec2}, the first image of the electrified case corresponds to the instant at which the voltage drop was established. Therefore, $R_{\textin{min}}$ approximately corresponds to that of the non-electrified case at that instant. This allows us to position the curves of the electrified cases in the graph.

In all the cases analyzed, the electric field significantly delayed the breakup. This delay explains some of the effects described above. Due to the strong asymmetry imposed by the disparity of the radii of the supporting disks, the first pinching point for $V_0=0$ is always located at the lower filament end. When the electric field is applied, most of the voltage drop accumulates in that region, which slows down the breakup process. If the electric field is sufficiently intense, the upper filament end ``catches up" the lower one and pinches almost simultaneously. This can be observed in the breakup of W70G30 for $V_0=400$ V and of W30G70 for $V=200$ V. In fact, this effect explains the formation of the satellite droplet from the electrified W30G70 filament (Fig.\ \ref{evol3}). As will be shown by the simulations, the pinching delay is caused by the polarization force. For this reason, this effect is less noticeable in the case of octanol, whose permittivity is much lower than that of the water-glycerine mixtures. In fact, it does not affect the evolution of the minimum radius for $R_{\textin{min}}\gtrsim 12$ $\mu$m. This explains why the electric field does not enhance the formation of the octanol satellite droplet. The maximum breakup delay observed in our experiments is around 60 $\mu$s. This value is much smaller than the liquid bridge breakup time, which is of the order of the inertio-capillary time $t_c\sim 10^4$ $\mu$s. Therefore, this phenomenon cannot be observed on a scale set by the breakup time.

The orange symbols for W70G30 and W30G70 and blue symbols for octanol correspond to experiments in which the same voltage was applied at two significantly different instants. Interestingly, the difference between the two corresponding curves $R_{\textin{min}}(t)$ is just a constant horizontal displacement. This means that the filament thinning rate, $dR_{\textin{min}}(t)/dt$, is essentially independent of the time at which the electric field was applied and only depends on the minimum radius $R_{\textin{min}}$. In other words, two electrified filaments with the same minimum radius thin at the same speed regardless of their ``electric history".

\begin{figure}
\begin{center}
\includegraphics[width=1\linewidth]{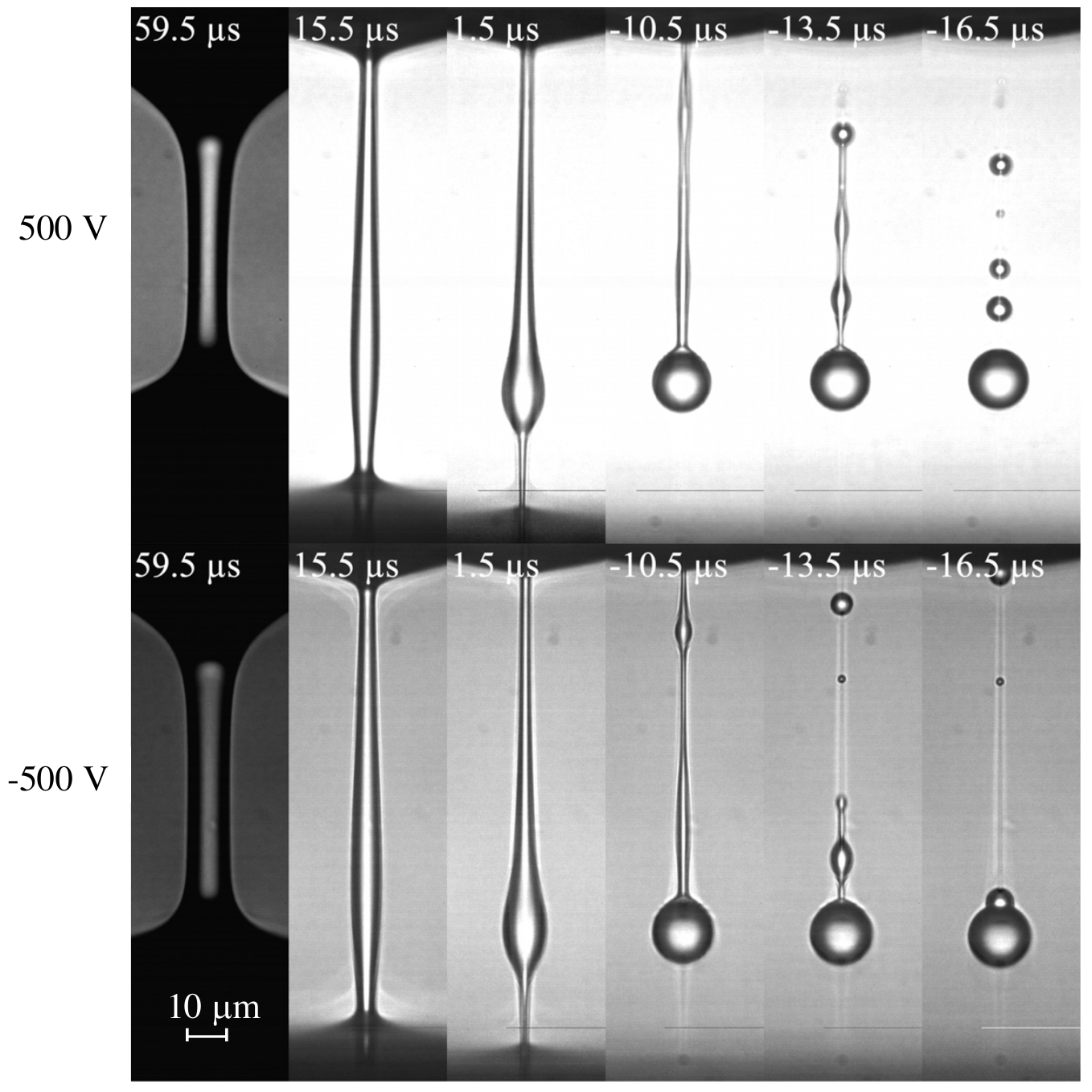}
\end{center}
\caption{Images showing the evolution of a liquid bridge of W44G56 for $V_0=500$ V with positive (upper images) and negative (lower images) polarities. The labels indicate the time to the pinching $\tau$. The voltage was applied at $\tau=61.5$ and 62.5 $\mu$s for the positive and negative polarity case, respectively.}
\label{evolp}
\end{figure}

The characteristic hydrodynamic time $t^*\equiv (R_{\textin{min}}^{-1} dR_{\textin{min}}/dt)^{-1}$ takes values of the order of 10 $\mu$s in the last phase of the breakup. This value is roughly of the same order of magnitude as that of the electric relaxation time $t_e$ (see Table \ref{tab1}). One may wonder whether electrokinetic effects affect the breakup dynamics in this case. Specifically, we explored the possible influence of the mobilities of positive and negative ions on the filament breakup. Figure \ref{evolp} compares images taken for the same experimental conditions but inverting the polarity. There are very slight differences between the two sequences of images, which may be attributed to the small difference between the instants at which the electric field was applied. We have verified that polarity has a negligible effect in the case of octanol too. Therefore, the difference between the mobilities of the positive and negative ions present in the liquid does not significantly affect the pinch-off dynamics.

\subsection{Numerical results}
\label{sec42}

This section analyzes numerically the breakup of W44G56 and octanol filaments to explain the experimental observations described in the previous section. The sets of dimensionless parameters in the simulation of W44G56 and octanol were $\{R=19$, $\Lambda=13.3$, $\hat{{\cal V}}=6574$, $B=2 \times 10^{-3}$, Oh=$0.099$, $\beta=69.7$, $\alpha=1971$, $\chi=0.056\}$ and $\{R=19$, $\Lambda=13.3$, $\hat{{\cal V}}=6747$, $B=3.8 \times 10^{-3}$, Oh=$0.16$, $\beta=10$, $\alpha=59.3$, $\chi=0.9\}$, respectively. The value of $\chi$
for W44G56 corresponds to $V_0=200$ V. We chose this case because the voltage $V_0=500$ V had to be applied very close to the pinching (Fig.\ \ref{Rmin0}) to allow the filament breakup. As will be explained below, this is related to the high liquid permittivity of W44G56. The value of $\chi$ for octanol corresponds to $V_0=500$ V.

\begin{figure}
\begin{center}
\includegraphics[width=0.7\linewidth]{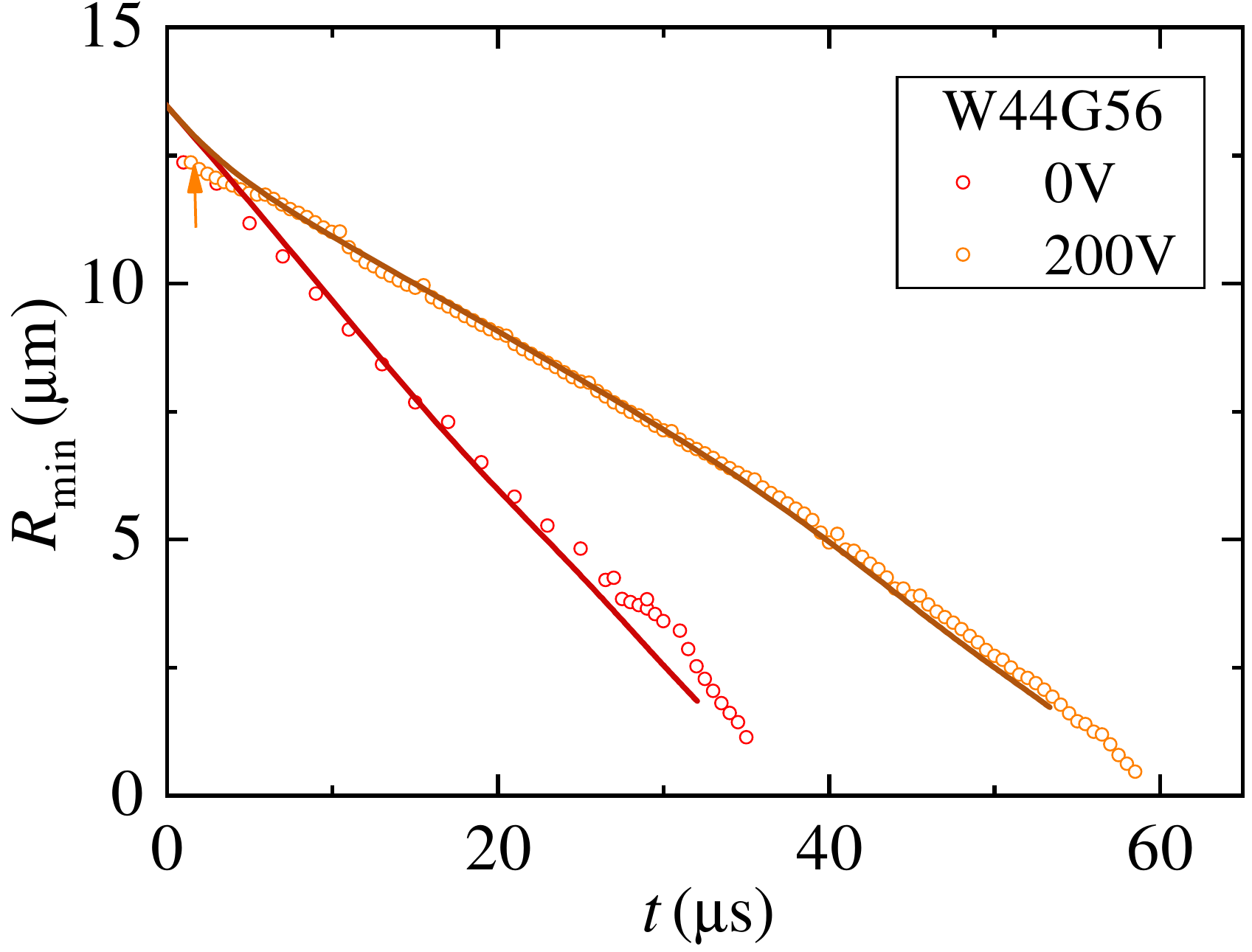}
\includegraphics[width=0.7\linewidth]{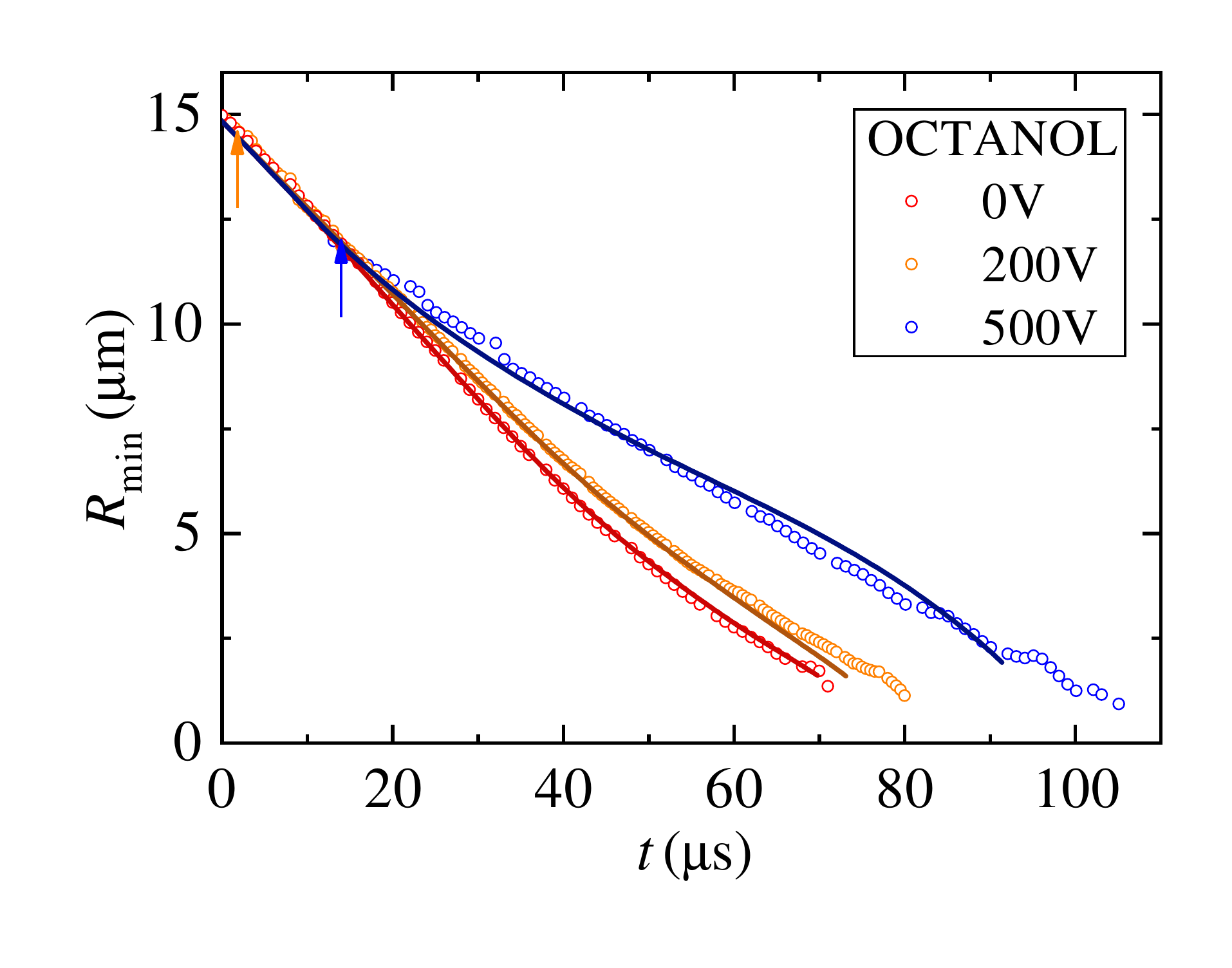}
\end{center}
\caption{$R_{\textin{min}}(t)$ for W44G56 and octanol. The symbols are the experimental data, while the solid lines correspond to the numerical simulations.}
\label{compara}
\end{figure}

\begin{figure}
\begin{center}
\includegraphics[width=\linewidth]{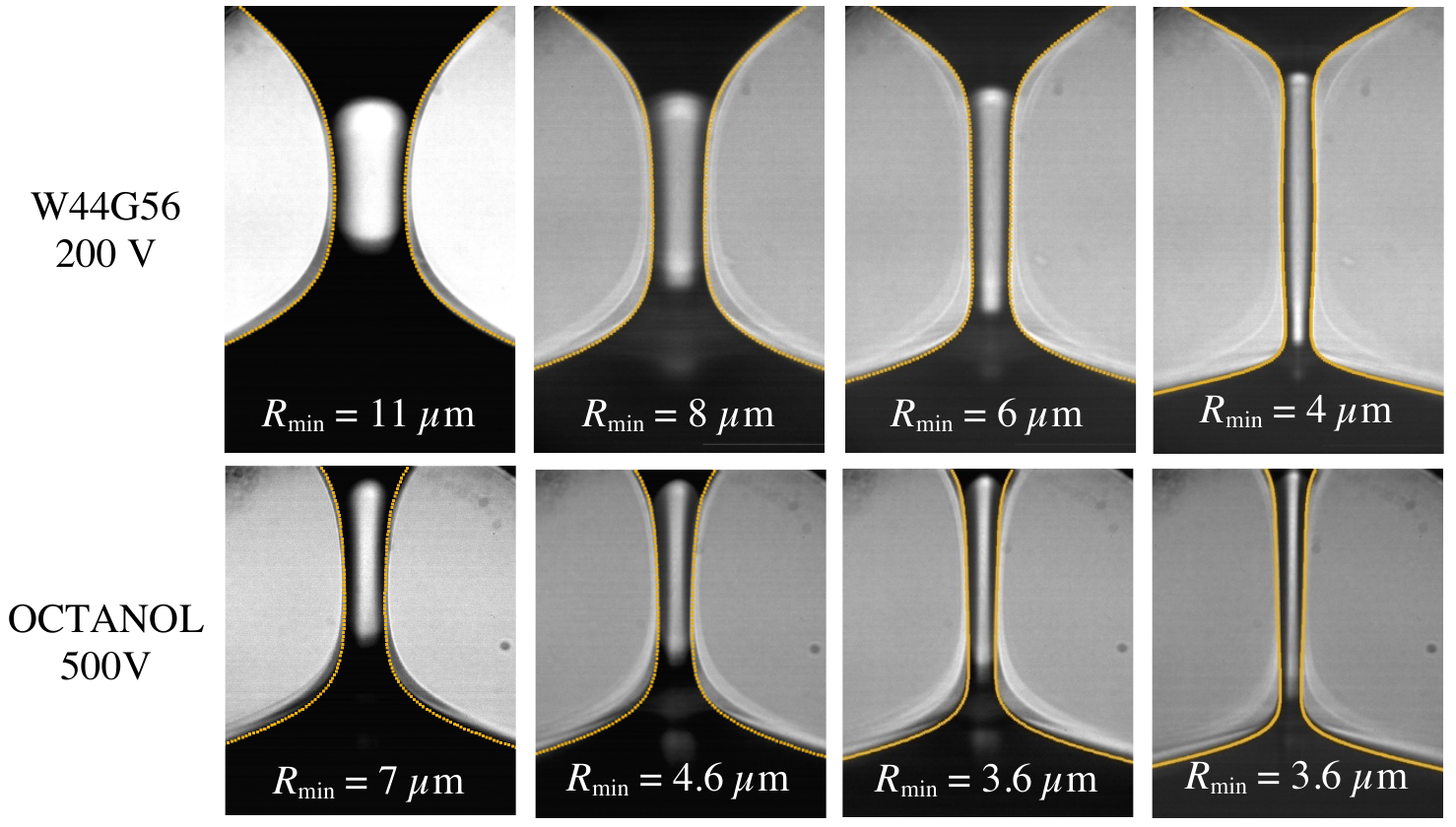}
\end{center}
\caption{Images of W44G56 for $V_0=200$ V and of octanol for $V_0=500$ V. The solid lines are the numerical contours calculated at the instants when $R_{\textin{min}}(t)$ took the same value as that of the experiment.}
\label{compara2}
\end{figure}

The solution of the leaky-dielectric model reproduced remarkably well the evolution of the electrified filaments when we slightly modified the values of upper triple contact line radius, liquid properties, and the instant at which the voltage was applied. This modification was smaller than the experimental uncertainty (around 10\%) in determining the values of those quantities. For the sake of illustration, Fig.\ \ref{compara} shows the agreement between the numerical and experimental results for $R_{\textin{min}}(t)$ during the thinning of W44G56 and octanol filaments. The shapes were almost the same (Fig.\ \ref{compara2}) even though the electrical boundary conditions in the experiment (Figs.\ \ref{sketch2} and \ref{evol}) differed from those of the simulation (Fig.\ \ref{sketchnum}).

\begin{figure}
\begin{center}
\includegraphics[width=0.7\linewidth]{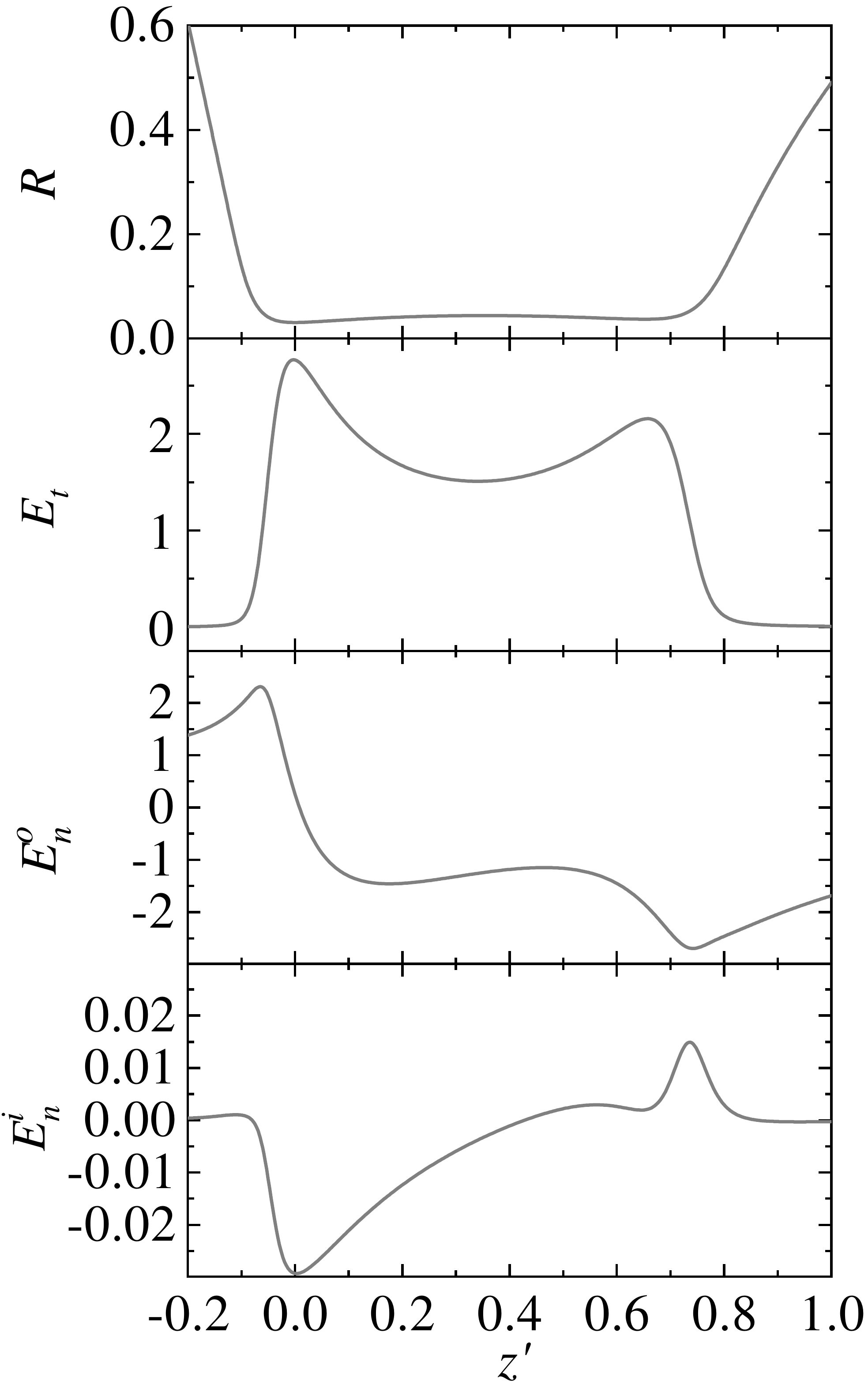}
\end{center}
\caption{Free surface contour $R$, tangential electric field $E_t$, outer normal electric field $E_n^o$, and inner electric field $E_n^i$ for W44G56 and $V_0=200$ V. The results calculated at the instants when $R_{\textin{min}}=0.0296$. All the quantities were made dimensionless as explained in Sec.\ \ref{sec3}.}
\label{field}
\end{figure}

Figure \ref{field} shows the tangential and normal components of the electric field along the W44G56 filament surface. The results are plotted versus the axial coordinate $z'=z-z_{\textin{min}}$, where $z_{\textin{min}}$ indicates the location of the minimum free surface radius. The interval $z'\lesssim -0.15$ corresponds to the lower parent drop. The normal outer component takes its higher values in that interval, while the inner component practically vanishes because the drop is quasi-equipotential. The same occurs in the upper parent drop ($z'\gtrsim 0.8$), although the sign of the normal outer component is negative because the droplet is in contact with the positive electrode. The condition $\beta E_n^i\ll E_n^o$ does not hold along the filament connecting the parent drops, implying that the electrical conductivity is not high enough for the surface charge to relax to its local electrostatic value and screen the applied electric field. In other words, the ratio between the electric relaxation time $t_e$ and the characteristic hydrodynamic time $t^*$ is not sufficiently small in this phase of the liquid bridge breakup for the surface charge to relax to the equipotential solution. The maximum value of the tangential electric field is similar to that of the outer normal electric field. However, $E_t$ reaches that value at the point with the minimum radius ($z'=0$), while $E_n^o$ practically vanishes at that point. This result allows us to anticipate that the polarization stress $\chi(\beta-1)(E_t)^2/2$ becomes dominant in the pinching region. 

\begin{figure}
\begin{center}
\includegraphics[width=0.7\linewidth]{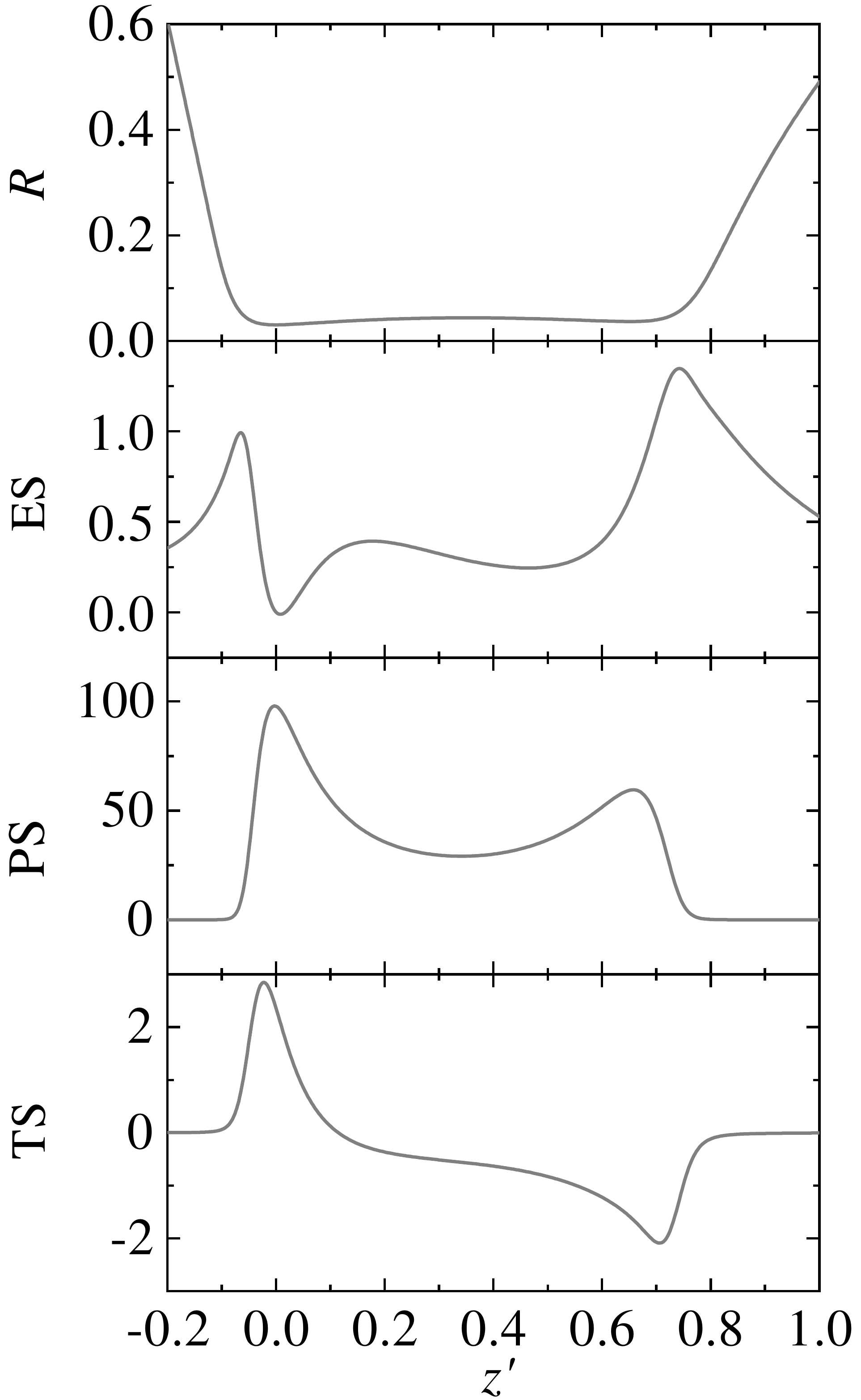}
\end{center}
\caption{Free surface contour $R$, electrostatic suction (ES), polarization stress (PS), and tangential stress (TS) for W44G56 and $V_0=200$ V. The results calculated at the instants when $R_{\textin{min}}=0.0296$. All the quantities were made dimensionless as explained in Sec.\ \ref{sec3}.}
\label{maxwell}
\end{figure}

\begin{figure}
\begin{center}
\includegraphics[width=0.8\linewidth]{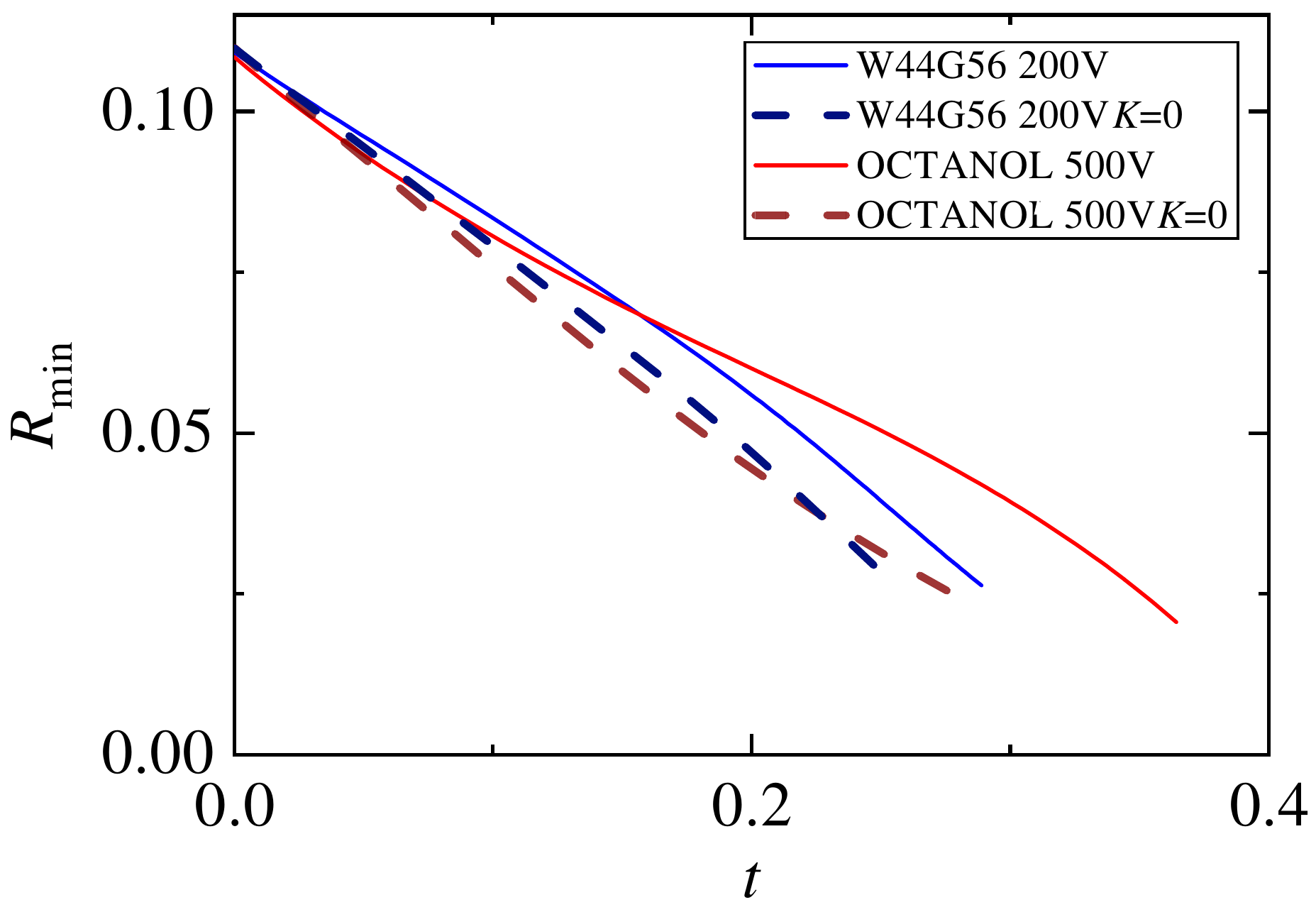}
\end{center}
\caption{$R_{\textin{min}}(t)$ for W44G56 and $V_0=200$ V (blue lines), and for octanol and $V_0=500$ V (red lines) with the true value of the electrical conductivity (solid lines) and $K=0$ (dashed lines). The origin of time $t=0$ corresponds to the instant at which the W44G56 and octanol filaments have the same minimum radius.}
\label{dielectric}
\end{figure}

\begin{figure}
\begin{center}
\includegraphics[width=0.75\linewidth]{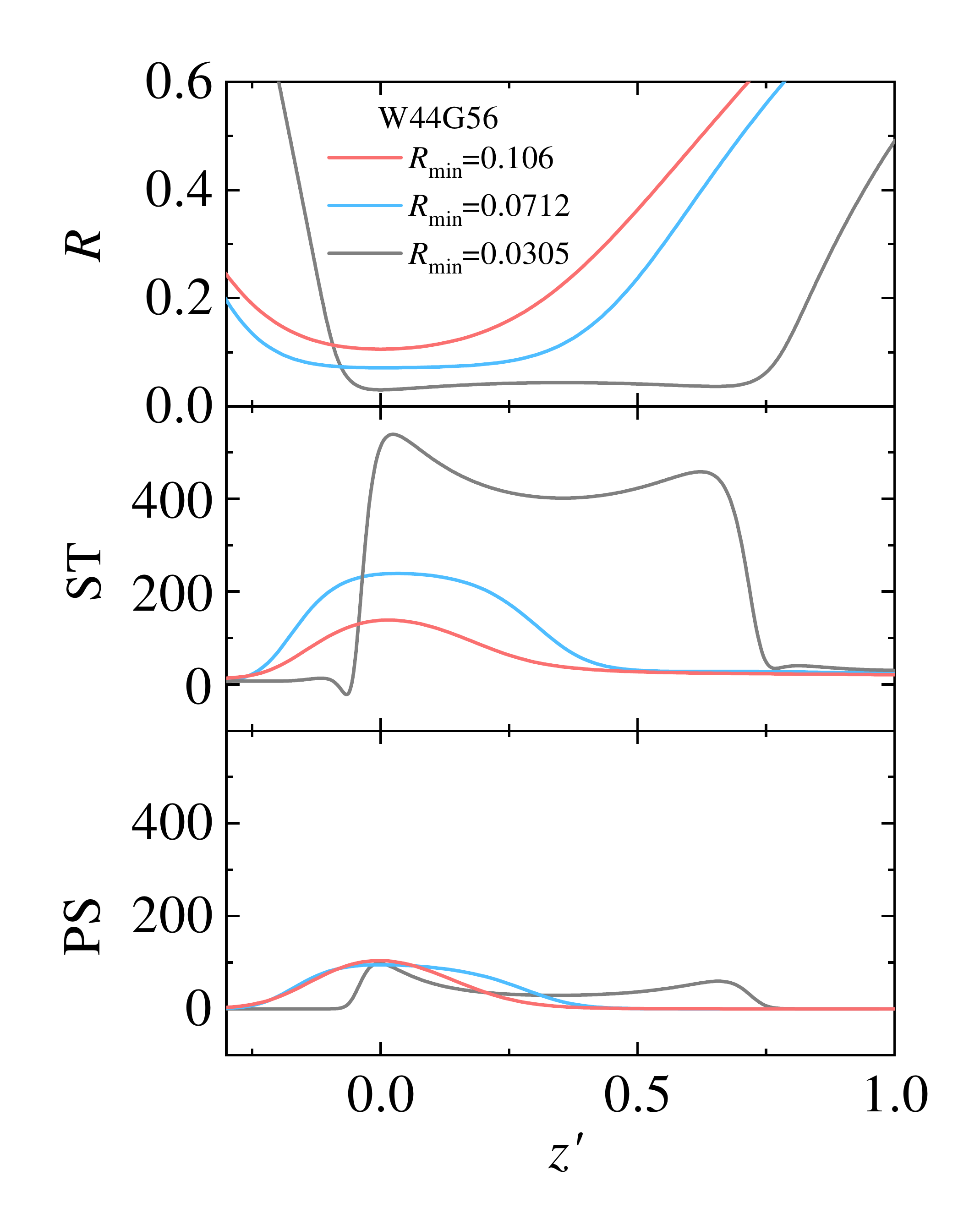}
\end{center}
\caption{Free surface contour $R$, surface tension stress (ST) and polarization stress (PS) for W44G56 and $V_0=200$ V. The results calculated when $R_{\textin{min}}$ took the three values indicated in the figure. All the quantities were made dimensionless as explained in Sec.\ \ref{sec3}.}
\label{maxwell2}
\end{figure}

\begin{figure}
\begin{center}
\includegraphics[width=0.7\linewidth]{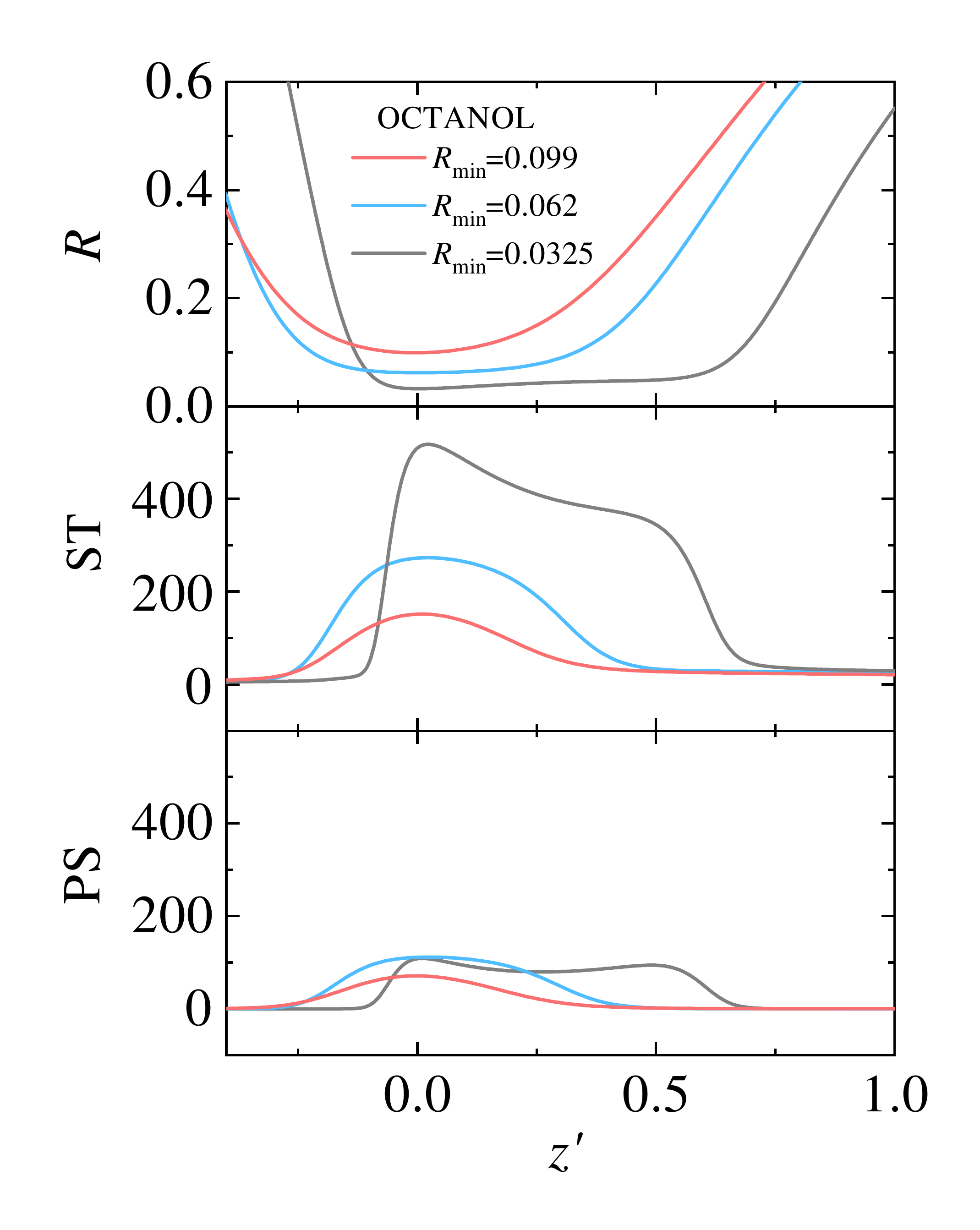}
\end{center}
\caption{Free surface contour $R$, surface tension stress (ST) and polarization stress (PS) for octanol and $V_0=500$ V. The results calculated when $R_{\textin{min}}$ took the three values indicated in the figure. All the quantities were made dimensionless as explained in Sec.\ \ref{sec3}.}
\label{maxwell3}
\end{figure}

Now, we examine the distribution of the capillary and Maxwell stresses along the liquid filament connecting the two parent drops. According to Eqs.\ (\ref{int3}) and (\ref{int2}), we define the surface tension (ST) stress, electrostatic suction (ES), polarization stress (PS), and tangential stress (TS) as
\begin{equation}
\text{ST}\equiv -\frac{RR_{zz}-1-R_z^{2}}{R(1+R_z^{2})^{3/2}}, \quad
\text{ES}\equiv\frac{\chi}{2}\left[(E_n^o)^2-\beta (E_n^i)^2\right], \end{equation}
\begin{equation}
\text{PS}\equiv\chi\frac{\beta-1}{2}(E_t)^2,\quad 
\text{TS}\equiv\sigma E_t.
\end{equation}
These stresses are plotted in Figs.\ \ref{maxwell}--\ref{maxwell3}. All the quantities were made dimensionless as explained in Sec.\ \ref{sec3}.

Figure \ref{maxwell} compares the three Maxwell stresses (ES, PS and TS) when $R_{\textin{min}}=0.0296$ in the W44G56 filament. As can be observed, the polarization stress PS takes values much higher than the electrostatic suction ES and the tangential stress TS. Due to the quasi-axial configuration of the electric field between the two parent drops, the charge accumulated on the free surface has negligible mechanical effects. The tangential electric field $E_t$ takes very high values in the filament, which explains the large polarization stress arising in that region.

As mentioned above, the polarization stress becomes much larger than those associated with surface charge. This does not imply that $R_{\textin{min}}(t)$ is the same as that of a dielectric liquid with the same permittivity because the electric field is different in the two cases. To show this, we conducted numerical simulations with zero electrical conductivity keeping the same values for the rest of the physical properties. As can be observed in Fig.\ \ref{dielectric}, $R_{\textin{min}}(t)$ is considerably different for the true value of $K$ and for the dielectric case $K=0$. The pinching delay in the leaky-dielectric case is larger than for $K=0$ because the voltage decay concentrates in the filament connecting the two equipotential parent drops.

Now, we analyze the competition between the driving surface tension and the polarization stress (Figs.\ \ref{maxwell2} and \ref{maxwell3}). ST and PS have the same sign. According to Eq.\ (\ref{int3}), this means that PS opposes ST and delays the pinching (Fig.\ \ref{compara}). This delay takes place essentially during the first phase of the filament thinning (cyan and red lines), when PS is commensurate with ST. In the time interval analyzed in Figs.\ \ref{maxwell2} and \ref{maxwell3}, the capillary pressure increases due to the thinning of the filament. The filament remains practically cylindrical, and its length increases. As a consequence, the axial electric field does not drastically increase next to the pinching point. For this reason, ST is much larger than PS in that region at the last instant considered ($R_{\textin{min}}=0.0296$). The polarization stress is expected to become subdominant in the rest of the pinching process, and surface tension must be balanced by viscosity. This tendency can be observed in Fig.\ \ref{compara}, which shows how $R_{\textin{min}}(t)$ seems to become ``parallel" to the curve for the non-electrified case. In other words, the rate of thinning is not affected by the electric field in the last phase of the pinching, even though the polarization stresses must diverge in the pinching point as the free surface approaches its breakup. Unfortunately, the simulation spatial and temporal resolutions did not allow us to reach smaller values of $R_{\textin{min}}$ to appreciate this behavior.

The dominance of the hydrodynamic forces in the ultimate phase of the free surface pinching is consistent with the scaling law for the non-electrified case. Figure \ref{Rmin} shows that, in the absence of the axial electric field, the pinching of the W44G56 interface is dominated by the viscous force in the last instants considered. According to the viscous scaling law (\ref{vis}), 
\begin{equation}
\label{vism}
\text{ST}\sim R_{\textin{min}}^{-1}\sim \tau^{-1}, \quad
\ell^{-2}\sim \tau^{-0.35}.
\end{equation}
To obtain an upper bound of $E_t$ in the pinching region, we assume that the applied voltage decays in that region asymptotically close to the pinching. In this case, 
\begin{equation}
\label{vism2}
\text{PS}\sim E_t^2\sim \ell^{-2}\sim \tau^{-0.35}.
\end{equation}
This implies that if the filament pinching enters into a ``hydrodynamic regime" in which the polarization stress is subdominant, it will remain in that regime even if the polarization stress diverges asymptotically close to the pinching point. In essence, viscosity stretches the pinching region (increases $\ell$), keeping the magnitude of polarization force below the hydrodynamic ones.

\section{Conclusions}
\label{sec5}

This paper analyzed experimentally and numerically the breakup of a Newtonian filament subject to an axial electric field. The experimental analysis showed the rich phenomenology arising when the electric field is applied. The most salient effects are:
\begin{itemize}
    \item The size of the W70G30 satellite droplets exhibits a non-monotonous dependency on the electric field magnitude. 
    \item The W44G56 filament detaches from the parent droplets and breaks up due to the electro-capillary instability. The surface charge distributes along the filament surface before its breakup, which produces satellite droplets with opposite charges.
    \item The electric field produces satellite droplets of W30G70, which do not form in the non-electrified case. This effect is not observed for octanol owing to its lower permittivity.
    \item  The electric force significantly delays the free surface pinching, which explains the formation of satellite droplets of the water-glycerine mixtures when the voltage is applied. 
    \item The breakup delay is much smaller than the liquid bridge breakup time. Therefore, this phenomenon cannot be observed on a scale set by the breakup time.
    \item Two electrified filaments with the same minimum radius thin at the same speed regardless of the instant when the voltage was applied.
\end{itemize}

The numerical simulations allowed us to explain some of the effects listed above. The major conclusions of this theoretical analysis are:
\begin{itemize}
    \item The solution of the leaky-dielectric model reproduced remarkably well the evolution of the electrified filaments.
    \item Although $E_t$ and $E_n^o$ take similar maximum values, $E_t$ reaches that value at the minimum radius, while $E_n^o$ practically vanishes at that point.
    \item The polarization stress becomes much larger than the rest of Maxwell stresses. However, the evolution of the minimum radius differs from that of the dielectric liquid with the same permittivity owing to the difference between the electric fields in the two cases.
    \item As the free surface approaches the pinch-off, the polarization stress becomes much smaller than the capillary one, and surface tension is balanced by viscosity. 
    \item The electric field does not affect the thinning rate asymptotically close to the pinching point, even though the polarization stress may diverge in that limit.
\end{itemize}

The leaky-dielectric model assumes that the electric relaxation time is much smaller than the characteristic hydrodynamic time. As the electrified filament approaches its breakup, this assumption fails, and charge relaxation phenomena become relevant \citep{LGPH15}. The comparison between our numerical simulations and experiments does not allow us to elucidate whether the transfer of charge from the bulk to the surface and the resulting surface electric field are correctly modeled because their dynamical effects are negligible.

Our analysis is restricted to moderately viscous liquids with high permittivity. The conclusions presented above may not apply to low-viscosity and/or low-permittivity liquids.

The maximum value of the tangential electric field in the simulation of W44G56 and octanol is around 0.4 MV/m and 1.2 MV/m, respectively. These values are much smaller than the corresponding characteristic value $E_o$ of electrospray (see Table \ref{tab1}). Despite this large difference, significant effects were observed in our experiments, which indicates the relevance of $E_t$ in the dynamics of jets produced with electrospray.

Finally, the search for an asymptotically self-similar pinching regime where electrical terms could play a non-subdominant role is an open problem. In the slender approximation, the electrical potential at the pinching can be expressed as $\phi^o=A(z,t) \ln (R(z,t))+B(z,t)$, which yields electric fields of the form $E_n^o=A/R$ and $E_t=-\phi^o_z\simeq -A R_z R^{-1}$. The complete formulation of the problem demands the consideration of the charge conservation equation, which in the slender approximation leads to
\begin{equation}
A_t+w A_z +A w_z/2=0.
\end{equation}
There is a specific time dependence of $A$ that renders the solution self-similar. Finding that solution and its spatiotemporal consistency, including its continuation to the far-boundary problem with specific features and time-dependent electrical connections, is a non-trivial task which will be the subject of future work.

\acknowledgments{This research has been supported by the Spanish Ministry of Economy, Industry and Competitiveness under Grants PID2019-108278RB, and by Junta de Extremadura under Grant GR18175.}


%

\end{document}